\documentclass[fleqn,usenatbib]{mnras}
\usepackage{subcaption}
\usepackage{newtxtext,newtxmath}

\usepackage[T1]{fontenc}
\usepackage{natbib}
\DeclareRobustCommand{\VAN}[3]{#2}
\let\VANthebibliography\thebibliography
\def\thebibliography{\DeclareRobustCommand{\VAN}[3]{##3}\VANthebibliography}

\usepackage{graphicx}	
\usepackage{amsmath}	

\title[Seeds of supermassive black holes]{Seeds of supermassive black holes in general relativistic and alternative cosmologies: Implications of massive seeds}

\author[]{
Das N.,$^{1}$\thanks{E-mail: nirmalidas995@gmail.com}
Kalita S.,$^{1}$
Kakati A.,$^{1}$
\\
$^{1}$Department of Physics, Gauhati University\\
}

\date{Accepted XXX. Received YYY; in original form ZZZ}

\pubyear{\the\year{}}

\begin{document}
\label{firstpage}
\pagerange{\pageref{firstpage}--\pageref{lastpage}}
\maketitle

\begin{abstract}

Presence of supermassive black holes (SMBHs) with mass $(10^{6}-10^{9}) M_{\odot}$ at $z = 10$ has been recently revealed by James Webb Space Telescope (JWST) observations. In this study we generate seeds for the above range of SMBHs in various background cosmologies. We consider cosmic timescales required for black hole growth provided by three general relativistic cosmological models ($\Lambda$CDM, $\omega$CDM and Dynamical Dark Energy(DDE) and the  braneworld  cosmology. The growth of SMBHs is studied through Eddington limited and super-Eddington accretion, where the accretion starts at z=30. It is found that growth of SMBHs by z=10 within Eddington limited accretion is possible through massive seeds $(M\geq10^{4}M_{\odot})$ in all cosmologies. Super Eddington accretion onto spinning black holes with mass of few tens of solar masses can result in SMBHs by z=10 in all cosmologies. The viable cosmologies considered here are found to be unable to strongly distinguish between the seed black hole masses. The seeds generated in this work are assumed to be of primordial origin in order to satisfy the criteria of formation of high redshift massive galaxies. The fraction of primordial black holes (PBHs) contributing to dark matter ($f_{PBH}$) and their corresponding number densities  for the mass range ($10^{5}-10^{8}$) $M_{\odot}$ are calculated in both seed effect and Poisson effect. In seed effect, PBHs of mass $\geq 10^{7} M_{\odot}$ contributes $\leq 10^{-2}$ to the dark matter fraction. The evolution of gas mass inside a PBH seeded dark matter halo is studied. The ratio of black hole to stellar mass is also evaluated for star formation efficiency in the range (0.1-1) and found to be ($10^{-3}-1$) for $M_{BH}=10^{8} M_{\odot}$ and ($10^{-2}-10$) for $M_{BH}=10^{9} M_{\odot}$.

\end{abstract}

\begin{keywords}
black hole physics -- quasars: supermassive black holes -- galaxies: high redshift
\end{keywords}



\section{Introduction}

Existence of supermassive black holes (SMBHs) in the centres of galaxies and active galactic nuclei (AGN) have been long-established. These black holes generally have mass range $(10^7-10^{9})M_\odot$ and are believed to have formed within less than 700 million years after the Big Bang. Black holes  as massive as $(10^8-10^{10})M_\odot$ have also been reported \citep{10.1093/mnras/stac2833}.These are believed to have formed 900 million years after the Big Bang.\\
Origin of massive black holes that power the quasars in the high redshift universe is a deep puzzle in astrophysical cosmology. It often challenges the usual astrophysics of black hole mass accretion or the timeline in standard model of cosmology. In the pre JWST era of observation, existence of a population of SMBHs of mass $(10^{9}-10^{10}) M_{\odot}$ at redshift z= 4-7.5 was recorded (\cite{fan2004black}; \cite{mortlock2011luminous}; \cite{banados2018800};  \cite{yang2021probing}). JWST has pushed the limit of high redshift to z = 4-11 for AGNs (\cite{greene2024uncover}; \cite{larson2023ceers}; \cite{andika2024tracing}; \cite{kocevski2024rise}). For spectroscopically confirmed AGNs the black hole masses are confirmed as $(10^{6}-10^{8}) M_{\odot}$ (\cite{harikane2023jwst}; \cite{maiolino2023massive}). Black hole mass to stellar mass ratio $(M_{BH}/M_{*})$ in the galaxies at z$\geq$ 4 indicate that overly massive black holes that are 10-100 times heavier than expected from local $(M_{BH}/M_{*})$ ratio exist in those high redshift galaxies (\cite{natarajan2023first}; \cite{kokorev2023uncover}). Even if these are exciting developments we are yet to understand the mechanism of generation of seeds of these very massive black holes.\\
Formation and growth of black holes have several channels. These are end products of evolutionary collapse of single or very massive stars,  accretion of mass onto seeds, runaway stellar mergers in young clusters or black hole merger, collapse of high-contrast density perturbations, collapse of gas clouds \citep{2021NatRP...3..732V}. The growth of SMBHs from accretion onto stellar mass black holes as seeds require a larger time span in order to power quasars by redshift 6-7, hence posing a timing challenge.  Baryonic matter accretion at a critical rate $f$=1, where $f$ is the Eddington fraction, requires massive seed of about $10^4 M_\odot$ to generate a $10^9 M_\odot$ SMBH within t=597 Myr \citep{10.1111/j.1365-2966.2006.10467.x}. Possible candidates for seeds include light seeds $(10^{2}-10^{3}) M_{\odot}$ of POP III stellar remnants and intermediate seeds $(10^{3}-10^{4}) M_{\odot}$ or Direct Collapse Black Holes (DCBHs) resulting from isothermal gravitational collapse of pristine gas clouds which fail to undergo fragmentation due to absence of metal cooling and molecular $H_2$ cooling (\cite{hbhowmick2024growth}). Accretion of self interacting dark matter (SIDM), of specific cross-section, onto seed black holes can produce massive black holes $(10^3-10^8) M_\odot$ which in turn can act as progenitors for SMBH formation \citep{das2024formation}. If massive black holes are in place in the pre-quasr epoch ($z\geq 6$) it challenges the cosmic timeline presented by standard model of cosmology. These massive black holes must grow within 500-900 Myrs after the Big Bang which is likely to be too short for growth- a problem identified as the 'time compression' \citep{melia2015supermassive}. Recent discovery of a high redshift (z = 6.3) quasar J0100+2802 inhabiting a central black hole of mass $\approx 10^9 M_\odot$ requires the super Eddington growth \citep{volonteri2005rapid} of the central SMBH within 900 million years after the Big Bang or the growth from massive seed black holes \citep{yoo2004formation} in the standard $\Lambda$CDM cosmology. If accretion has to occur at or within the Eddington rate existence of very massive seeds at high redshifts is a necessary condition. This may be in tension with the standard $\Lambda$CDM cosmology, particularly due to lack of adequate number of halo sites for these objects (see \cite{suh2024super} and references therein). The tension is likely to get reduced by invoking super-Eddington accretion of lighter seed black holes (\cite{volonteri2015case}; \cite{bramberger2015cosmic} ). From  the deep JWST/NIRSpec observations of an AGN, Abell2744-QS01 at redshift z = 7.0541, the central black hole mass is found as $10^7 M_\odot$. It is accreting at the rate of of 30\% $L_{Edd}$, where $L_{Edd}$ is the Eddington luminosity \citep{furtak2024high}. JWST report of an ultra luminous galaxy GN-z11 at z=10.6 reveals the transitions of a typical AGN hosting a black hole of mass $M_{BH}=10^{6.2\pm0.3} M_{\odot}$ accreting at a rate of about $5L_{Edd}$ \citep{maiolino2024small}. Cosmological simulations show that Eddington accretion onto heavy seeds and super-Eddington  accretion onto light or intermediate seeds can reproduce the mass of GN-z11 at z = 10.6. \cite{zhang2023super} demonstrated that starting with accretion at sub-Eddington rate and intermittently at super-Eddington onto an intermediate mass seed ($10^3 M_\odot$) at z = 15, can explain the mass of GN-z11 and black hole to the stellar mass ratio.	The discovery of luminous high redshift quasars at $z>6$ \citep{Bañados_2014,mortlock2011luminous}, comprising central black hole of mass $10^9 M_{\odot}$, poses the problem of timeline predicted by the standard $\Lambda$CDM model. The general problem is that we do not understand how these massive black holes originate in the infancy of the universe.\\
Cosmological models have an important role in understanding timeline of black hole formation. The evolution of the massive structures like black holes involves serious attention to the cosmology of the early universe. Primordial black holes (PBHs) formed in the early universe from the primordial inhomogeneities are potential candidates of massive black hole progenitors. The formation of PBHs from large inhomogeneities might have been enhanced either by the effects of the initial conditions of the universe or some kind of phase transition \citep{carr2021constraints}. Initial quantum effect during inflation is one possible source of fluctuations that could generate PBHs \citep{ivanov1994inflation}. A period of soft equation of state might have determined the mass range across which the PBHs form for a particular spectrum of primordial perturbations \citep{carr2021constraints}.\\
If PBHs form due to large density inhomogeneities in the very early universe, their mass depends on cosmic time (t) as M$\sim c^{3}t/G$. Lunar mass PBHs form during $10^{-14}$ sec after the Big Bang. Asteroid mass PBHs form during ($10^{-22}-10^{-17}$) sec. If LIGO black holes ($\sim 30-100) M_{\odot}$ are PBHs, they formed ($10^{-5}-10^{-4}$) sec. A PBH forming at t = 1 sec, the epoch of neutron-proton freeze out carries a mass of $10^{5} M_{\odot}$.\\
PBHs can provide the non baryonic dark matter \citep{carr2018primordial}. They are astrophysical Cold Dark Matter (CDM) candidates without requirement of new physics \citep{chapline2016new}. There are few mass windows of PBHs which can contribute significantly to the dark matter mass budget. The PBH dark matter mass function is represented by $f_{PBH}=\rho_{PBH}/\rho_{CDM}$, ($\rho_{CDM}$ being the total mass density of CDM)  \citep{carr2010new, carr2020primordial}. Lunar ($10^{20}-10^{24}$)g and asteroid ($10^{16}-10^{17}$)g mass PBHs and those of mass range similar to the LIGOs black holes ($\sim$10-100) $M_{\odot}$ are expected windows for a significant $f_{PBH}$ (close to unity).\\
The recent discoveries of massive black holes ($10^{6.2}-10^{7.9}) M_{\odot}$ at z = 10.3-10.6 \citep{2020ARA&A..58...27I, 2023ARA&A..61..373F} and massive galaxies with stellar mass $\ge 10^9~ M_{\odot}$ at z$\ge$ 10 \citep{Atek, Finkelstein,2023ApJS..265....5H,2022arXiv220802794N,2023ApJ...942L...9Y} have been sufficiently motivating for considering massive PBHs and their connection with galaxy formation scenario. Their existence leads to the speculation that either these black holes are themselves primordial in nature or they require rapid seeding process in the early universe. In a recent observation \citep{bogdan2024evidence} reported existence of a 5$\times10^{7} M_{\odot}$ at z $\geq$ 9 in an object to the class of Little Red Dot (LRD) discovered by JWST. This object is found to have $M_{BH}/M_{*} > 2$ thereby suggesting first detection of a PBH which is yet to experience the episode of accretion. It was argued earlier that massive PBHs can grow by accretion and form the SMBHs \citep{bean2002could}.\\
Massive PBHs can serve other purpose. They can seed galaxies through the seed effect and Poisson density fluctuations \citep{carr2018primordial, meszaros1975primeval}. PBHs being dark matter and massive PBHs providing seed for galaxies and massive black holes can be uncorrelated. However, as argued by \cite{carr2018primordial}, if PBHs have wide mass spectrum, they can serve both the purposes. Therefore, generating massive PBH seeds is interesting for addressing these two important problems in cosmology.\\
The idea of a universe loitering (dip of the Hubble parameter) at early redshift can allow a larger time span to boost the growth of gravitationally bound systems such as black holes as massive as $10^{9} M_{\odot}$ \citep{sahni2005did}. The universe in modified gravity models such as 5D braneworld can loiter at redshifts 6 $\leq$ z $\leq 40$, where the Hubble parameter, compared to $\Lambda$CDM, decreases during the loitering epoch and initiates the growth of high redshift black holes \citep{sahni2005cosmological}.\\
Formation of massive black holes also requires attention to the cosmogony involving nature of dark matter. If dark matter had sufficient self-interactions at extreme high redshifts, it can collapse to form black holes. Gravothermal collapse of self-interacting dark matter (SIDM) halo, due to negative specific heat, is eligible to produce high redshift massive black holes \citep{balberg2002gravothermal}. SIDM with specific cross section (0.5-5) $cm^2/g$ can gravothermally collapse to generate black holes as massive as $(10^{3}-10^{9})M_{\odot}$ \citep{das2024formation}.\\
Another alternative progenitor for SMBHs are cosmic string loops \citep{bramberger2015cosmic}. The intersection of cosmic strings can eventually collapse into black holes if they accumulate enough dark matter to cause collapse or if the loops have sufficient angular momentum to contract within the Schwarzschild radius \citep{hawking1989black}.\\
In this work we investigate seed masses required to form massive black holes by the epoch z = 10. We consider spinning black hole seeds in super-Eddington accretion. Both super Eddington and Eddington accretion are taken into account to generate massive black holes in timeline presented by different cosmological models. We consider three cosmological models within general relativity (GR) namely $\Lambda$CDM, $\omega$CDM and dynamical dark energy (DDE), with Chevalier, Polarski and Linder (CPL) parametrization \citep{chevallier2001accelerating, linder2003exploring} of equation of state and one cosmology beyond GR- the flat braneworld model. The cosmological models affect the expansion rate, thereby changing the time gap between any two epochs, which enter into the estimation of seed black hole mass. After generating the seed masses we assume that these seed black holes are of primordial origin and then obtain their dark matter fraction required to satisfy the criteria for massive galaxy formation with seed and Poisson effect. We also estimate the black hole to stellar mass ratio  for our seed black holes in all the cosmologies.\\
The manuscript is organized as follows. In section 2 we discuss accretion scenario of black hole growth and present timescale provided by background cosmological models. In section 3 we present seed masses required to form massive black holes. In section 4 we discuss the implications of the seed black holes obtained. Section 5 presents results and discussion. We conclude the main results in section 6.\\

\section{Black hole growth and timescale}
\subsection{Black hole growth by accretion}

Black hole mass growth and accretion rate are related via the accretion efficiency $\epsilon$ as
\begin{equation}
	\dot{M}_{BH} = (1-\epsilon)\dot{M}_{acc}
\end{equation}
Here $M_{BH}$ is mass of the black hole.
The accretion rate of black holes is expressed as
\begin{equation}
	\dot{M}_{acc}=\frac{L}{c^2}
\end{equation}
The luminosity, L is expressed in terms of the Eddington limit ($L_{Edd}$) as L=$fL_{Edd}$ where $f$ is known as the Eddington fraction. Here $L_{Edd}= M_{BH} c^2/t_E$, with $t_{E}= e^4/(24 \pi^2 \epsilon^{2}_o m^{2}_{e} c^{3} G m_{p})$ being Salpeter-Eddington time scale which is 469 Myrs. With this the black hole mass grows according to\\	 
\begin{equation}
	\dot{M}_{BH}= \left(\frac{f}{\epsilon}\right) \left(\frac{M_{BH}}{t_E}\right)  (1-\epsilon)
\end{equation}
The solution of above equation gives the seed mass $M_s$ which grows to a black hole of mass $M_{BH}$ at redshift $z_f$\\	
\begin{equation}
	M_s = M_{BH}(t(z_f)) exp\left(\frac{-f(1-\epsilon)}{\epsilon}\frac{t(z_f)-t(z_i)}{t_E}\right)
\end{equation}	
Here, $z_{i}$ is the initial epoch of accretion ($z_{i}>z_{f}$). We adopt an ideal assumption of uniform accretion throughout the time during which the seed mass evolves i.e. the duty cycle being close to unity. It means that the accretion rate of the seed black hole is always close to the mass transfer rate (duty cycle being the ratio of mass transfer to accretion rate). Therefore, the seed mass to be estimated in this study would represent the lower limit. In realistic case duty cycle may become smaller than unity, representing lesser time for activity of the seed accretion (accretion rate exceeding the mass transfer rate). It gives rise to massive seed mass.\\
Black hole mass growth is also affected by black hole spin ($\chi$). A fitting function for Eddington fraction exists in literature \citep{massonneau2023super} which is expressed as\\
\begin{equation}
	f=A(\chi)\left[\frac{0.985}{\frac{1.6\dot{M}_{Edd}}{\dot{M}_{acc}}+B(\chi)}+\frac{0.015}{\frac{1.6\dot{M}_{Edd}}{\dot{M}_{acc}}+C(\chi)}\right]
\end{equation}
Here
\begin{equation}
	A(\chi) = (0.9663-0.9292\chi)^{-0.5639}
\end{equation}
\begin{equation}
	B(\chi) = (4.627-4.445\chi)^{-0.5524}
\end{equation}
\begin{equation}
	C(\chi) = (827.3-718.1\chi)^{-0.7060}
\end{equation}
Here, $\dot{M}_{Edd}=L_{Edd}/\epsilon c^2$, with $\epsilon$ = 0.1. For Eddington limited accretion, we have the accretion ratio $a=\dot{M}_{acc}/\dot{M}_{Edd}$ = 1. In any other model this ratio departs from unity. The time scale allotted for growth of the seed, $\Delta t= t(z_f)-t(z_i)$, depends on expansion rate of the universe and hence the background cosmological model. For each cosmological model the seed mass required to form a final massive black hole at a given redshift depends on the Eddington fraction which in turn is governed by the black hole spin. $f<1, =1, >1$ respectively denote the sub-Eddington, Eddington and super-Eddington accretion.\\
Physics of super-Eddington accretion involves advections of photons liberated in the process of mass accretion onto the seed black hole. One of the important features of this regime is reduction of accretion efficiency in advection due to inability of photons to diffuse outward. \cite{madau2014super} showed that with advection induced reduction of accretion efficiency (efficiency below 0.1) a 100 $M_{\odot}$ seed can grow into $10^{9} M_{\odot}$ black hole within 1 Gyr. With JWST, \cite{suh2024super} has recently reported observational evidence of super-Eddington growth of a low mass $(7.2 \times 10^6 M_{\odot})$ black hole at z $\approx$ 4. Non-sphericity of accretion causes violation of Eddington limit of accretion (\cite{brightman2019breaking}). \cite{aird2019x} reported that such condition seems to be available in the pre-reionization era (The reionization epoch for $\Lambda$CDM is $z_{reion} \sim$ 7) \citep{montero2024five}. Super-Eddington accretion may shed new light in the scenario of massive black hole seeding with standard seeds expected from death of POP III stars, without invoking unusually massive seed black holes.\\

The growth of SMBHs and co-evolution with their host spheroids can result from phases of super-Eddington growth, coupled with star formation and positive feedback \citep{volonteri2015case}. However, recent JWST observations perhaps call for heavy seed black holes. Therefore, it is interesting to see how light or heavy a black hole seed can be in super Eddington accretion.\\

\subsection{Background cosmology and timescale} \label{sec:floats}
	The time interval $\Delta t\left(= t(z_f)-t(z_i)\right)$ in equation (4) is governed by the  expansion history H(z) in a cosmological model with H(z) being the time evolution of the Hubble parameter H,\\
\begin{equation}
	\Delta t= t(z_f)-t(z_i)= \int_{z_f}^{z_i} \frac{d z^{\prime}}{(1+z^{\prime})H(z^{\prime})}
\end{equation}
Here $t(z_i)$ is the initial epoch corresponding to redshift $z_i$ and the expansion history at a given epoch $z^{\prime}$ is expressed as \\	
\begin{equation}
	H(z^{\prime})= H_o f(\Omega_{i}, \omega_{i}, z^{\prime})
\end{equation}
where $\Omega_{i}$ are cosmological mass density parameters, $\omega_{i}$ is the equation of state (EoS) for the $i^{th}$ energy component and $z^{\prime}$ is the cosmological redshift.\\
In the quest for available timescales for the growth of massive black holes, we consider three different cosmological models within GR namely $\Lambda$CDM, $\omega$CDM with a constant equation of state and dynamical dark energy(DDE) with a varying equation of state in  Chevalier, Polarski and Linder (CPL) parametrization \citep{chevallier2001accelerating}. Outside GR we consider the spatially flat five dimensional braneworld cosmology \citep{dvali20004d}.

\subsubsection{Hubble parameter for GR based cosmologies}
The form of Hubble parameter H(z) is dependent on cosmological models. In spatially flat $\Lambda$CDM cosmology the H(z) function takes the form\\
\begin{equation}
	H(z)=H_{0}\sqrt{\Omega_{m(0)}(1+z)^{3}+\Omega_{\Lambda}},
\end{equation}
where $\Omega_{m(0)}$ is the current value of non-relativistic matter density parameter and $\Omega_{\Lambda}$ is the cosmological constant density parameter. The model parameters are $H_{0}$ = 67.4 kms$^{-1}$Mpc$^{-1}$, $\Omega_{m}=0.315$, $\Omega_{\Lambda}=0.6847$ \citep{aghanim2020planck}.
In $\omega$CDM cosmology, for a constant EoS $(\omega)$ of dark energy, the form of H(z) is given by\\
\begin{equation}
	H(z)=H_{0}\sqrt{\Omega_{m}(1+z)^{3}+\Omega_{DE}(1+z)^{3(1+\omega)}}
\end{equation}
From the Dark Energy Survey (DES)-SN5YR  fit to the flat $\omega$CDM model has given parameters of the model as $\Omega_{m}=0.264^{+0.074}_{-0.096}$, $\Omega_{DE}$ = 1- $\Omega_{m}$= 0.736, $\omega=-0.80$ \citep{abbott2024dark}. The value, $H_0$ = 71.1 kms$^{-1}$Mpc$^{-1}$ in $\omega$CDM is taken as the average of the available measurements of $H_0$ from various observations (\cite{riess2019large};\cite{chen2019sharp}; \cite{dominguez2019new};\cite{hotokezaka2019hubble}; \cite{freedman2019carnegie}; \cite{bonvin2017h0licow}; \cite{shajib2020strides}; \cite{hinshaw2013nine}; \cite{aghanim2020planck}.\\
Although $\Lambda$CDM has been in good agreement with earlier observations in cosmology, several recent and independent observations have brought challenges to the simple model. The most significant tension is the  $H_{0}$ tension, a statistically significant discrepancy of around 5.3$\sigma$ reported from CMB measurements and late time expansion rate measurements \citep{riess2022cluster}. A novel alternative within the framework of GR is that dark energy may be dynamical in nature. The simple dynamical dark energy (DDE) is characterized by time evolving equation of state with two parameters. It is known as CPL parametrization of dark energy and the EOS is written as\\
	\begin{equation}
		\omega_{DE}(a)= \omega_{0}+\omega_{a}(1-a) ,
	\end{equation}
where $\omega_{0}$ is the present day value of the EOS, $\omega_{a}$ is the derivative of the EOS with respect to the scale factor $a$.
For time varying EOS, in CPL parametrization, the Hubble parameter is \citep{ruiz2010dark}\\
	\begin{equation}
		H(z)=H_{0}\sqrt{\Omega_{m}(1+z)^{3}+\Omega_{DE}f(z)}\\
	\end{equation}
	where, 
	\begin{equation}
		f(z)=(1+z)^{(1+\omega_{0}+\omega_{a})}exp(-3\omega_{a}\frac{z}{1+z})
\end{equation}
Fitting DES-SN5YR to the flat DDE (CPL) model gives $\Omega_{m}=0.495^{+0.033}_{-0.043}$ , $\Omega_{DE}=1-\Omega_{m}=0.505$, $\omega_{0}=-0.36^{+0.36}_{-0.30}$ and $\omega_{a}=-8.8^{+3.7}_{-4.5}$ \citep{abbott2024dark}. The value of $H_{0}$ is taken to be 71 kms$^{-1}$Mpc$^{-1}$ from various observations. Recent DESI 2024 data release 1 (DR1) has considered CPL parametrization and reported preference of more than 2$\sigma$ level for DDE over a cosmological constant  \citep{lodha2025desi}.This measurement has been reported only through high redshift type Ia supernovae (SNeIa only) of full 5 year data set of DESI. The evidence for DDE has been found to be stronger by joint measurement Planck+DESI+DESY5 (henceforth written as PDDS) which has given $\omega_{0}=-0.726\pm0.069$, $\omega_{a}=-1.05^{+0.35}_{-0.28}$, $\Omega_{m}=0.3161$ and $H_{0}=67.22\pm0.22$ kms$^{-1}$Mpc$^{-1}$ \citep{giare2024robust}. It is compatible with present day EoS ($\omega_{0}$) lying within quintessence domain ($-1<\omega_{0}<-1/3$). Value of $\omega_{a}$ signifies phantom like behavior of dark energy at high redshift. DESI data has undergone significant improvement in the parameter space ($\omega_{0},\omega_{a}$).
\\
In this study we use both DDE (SNeIa only) \citep{abbott2024dark} and DDE (PDDS) \citep{giare2024robust}  constraints on the EoS.\\

\subsubsection{Hubble parameter outside GR: Braneworld models}
Higher dimensional cosmological models are promising modifications of GR. Motivated by string theories these models modify cosmological expansion dynamics. The very first higher dimensional cosmology was the Randall-Sundrum model which is a 5 dimensional gravitation theory \citep{randall1999large}. A 4 dimensional FLRW brane, where matter and radiation are confined is embedded in a 5 dimensional Anti de Sitter (AdS) bulk. This model predicts leakage of gravity to the higher dimensional bulk at extremely high energy scale, thereby altering cosmic expansion history immediately after the Big Bang. The curvature length scale of the AdS bulk has been recently constrained by black hole shadow based observations \citep{lemos2024hunting}. Formation of stable black holes in Randall-Sundrum Braneworld models was investigated by (\cite{chamblin2000brane}, \cite{biggs2022rotating}). On larger cosmological scale (hence in low energy regime) a 5 dimensional gravitation theory with  a 4 dimensional FLRW brane embedded in a 5 dimensional Minkowski bulk was formulated by Dvali, Gabadadze and Porrati \citep{dvali20004d}. The theory was extended by \cite{dvali20004d} to cosmology. Here leakage of gravity occurs once the universe expands beyond certain scale, known as the crossover scale. The consequence of this leakage is self accelerating universe which takes into account the late time cosmic acceleration without introducing negative pressure sources (see for example \cite{deffayet2002accelerated}).\\ It is possible to explain formation of massive structures in the early universe by loitering cosmology. During loitering phase the universe enters into a phase of minimum expansion, known as “loitering” (hesitating). The Hubble parameter is a minimum at this phase. In this nearly static phase the density perturbations in matter grows exponentially leading to burst of formation of gravitationally bound structures. Existence of a loitering epoch in the cosmic dawn (high redshift or prior to the epoch of cosmic reionization) was advocated by \cite{sahni2005did} to stretch the cosmic timescale (minimum Hubble parameter), thereby to allow not only formation of massive black holes at high redshift ($z\geq6$) but also existence of POP III  stars at z $\geq$ 10. \cite{sahni2005did} showed that it is a formidable challenge to produce loitering in general relativistic cosmology without adding unnaturally large spatial curvature which is not in conformity with inflationary paradigm and observation of CMB anisotropies. They successfully generated a loitering phase in a spatially flat braneworld cosmology with cosmological density parameters contributed by matter ($\Omega_{m}$), brane tension ($\Omega_{\sigma}$) dark radiation ($\Omega_C$) (projection of higher dimensional degree of freedom onto the brane), crossover scale ($\Omega_{l}$) and bulk cosmological constant ($\Omega_{\Lambda_{ b}}$). The loitering phase was generated around z = 18 and it was reported that it is possible to form very massive black holes within cosmological redshift z = 17-18. They found significantly larger age in the high z universe relative to the age expected in the standard $\Lambda$CDM cosmology. However, their remark lacked exact quantitative account of possible mass scales of black holes that could form by realistic accretion scenarios for differently spinning seed black holes. In addition to this, the model was scrutinized by \citep{elgaroy2005loitering} for consistency with CMB measurements. Any shift in the Hubble parameter and hence the cosmic age causes a shift in the peaks of the CMB fluctuation spectrum which is encoded in the CMB shift parameter R=$\sqrt{\Omega_{m}}H_{0}r(z_{dec})$, where r($z_{dec}$) is the commoving distance at decoupling redshift. Therefore, CMB shift parameter constrains ranges of the braneworld cosmological parameters \citep{sahni2005cosmological}. In order to generate a realistic braneworld cosmology at high redshift. \cite{elgaroy2005loitering} adopted the expansion rate equation of Sahni and Shtanov given by 
\begin{multline}
	H^{2}(z) = H_{0}^{2}( \Omega_{m}(1+z)^{3}+\Omega_{\sigma}+2\Omega_{l}\\-2\sqrt{\Omega_{l}}[\Omega_{\sigma}+\Omega_{l}+\Omega_{m}(1+z)^{3}+\Omega_{\Lambda_{ b}}+\Omega_{C}(1+z)^{4}]^{1/2})
	\end{multline}
The density parameters are given as (in the unit c=1)
$\Omega_{m}=\rho^{0}_{m}/(3m^{2}H^{2}_{0}),\  \Omega_{\sigma}=\sigma/(3m^{2}H_{0}^{2}),\   \Omega_{l}=1/(l^{2}H^{2}_{0}),\ \Omega_{\Lambda_{ b}}=-\Lambda_{b}/(6H^{2}_{0}),\ \Omega_{C}=C/(a^{4}_{0}H^{2}_{0}$)
Here, $1/m^{2}= 8\pi G$ and $l$ is the crossover scale. $\Lambda_{b}$ is the bulk cosmological constant and C is dark radiation density. At very high redshift the expansion rate reduces to \citep{elgaroy2005loitering}
\begin{equation}
	H^{2}(z)=H_{0}^{2}( \Omega_{m}(1+z)^{3}+2\sqrt{\Omega_{l}\Omega_{\Lambda_{ b}}}-2\sqrt{\Omega_{l}}[\Omega_{\Lambda_{ b}}+\Omega_{C}(1+z)^{4}]^{1/2})
	\end{equation}
In the present study we take density parameters which are compatible with CMB shift parameter given by WMAP observations \citep{spergel2003first}. The parameters are $\Omega_{m}$ = 0.3, $\Omega_{l}$ = 0.3, $\Omega_{C}$ = 1 and $\Omega_{\Lambda_{ b}}=10^{5}$ \citep{elgaroy2005loitering}. \cite{elgaroy2005loitering} reported that CMB measurements are respected by this parameter set. The value of H$_{0}$ is taken as 69.08$^{+0.71}_{-0.70}$ kms$^{-1}$Mpc$^{-1}$ \citep{bag2021phantom}. They produce only mild loitering (increase in cosmic age is only within 6$\%$ of the  $\Lambda$CDM age). We use these parameters in the braneworld Hubble function to estimate the seed black hole masses.\\
The Hubble function for all the cosmological models are displayed in Figure 2. . The fractional deviation of cosmic age from $\Lambda$CDM value at different redshifts is displayed in Figure 3. For the braneworld model the Hubble function is plotted with the high redshift approximation (equation 17). We see that except DDE cosmology constrained by SNeIa only, all cosmologies including DDE constrained jointly by Planck, DESI and DESY5 show smooth evolution of expansion rate. This model produces sufficiently large expansion rate at all redshifts, thereby reducing cosmic ages (see Figure 3). Our goal is to investigate models which stretch cosmic age at high redshift, thereby allowing sufficient growth time for accreting black holes. Although DDE (SNeIa only) model has undergone significant improvement in its parameter space we have kept it to see maximal effect of such models on the seed black hole mass required to form massive black holes in the pre-quasar eras (z$>>$7).\\
From calculation of ages in these cosmological models we see that except DDE (SNeIa only) all alternative models including DDE (PDDS) increases cosmic age relative to the benchmark cosmology at all redshifts (Figure 3). Braneworld cosmology significantly increases age at sufficiently low redshift. However, this increment softens at high redshift. At z = 0.8 and above, the cosmic ages increases to less than or equal to 6$\%$ of the $\Lambda$CDM. The fractional increment of age reduces to about 0.04
at z=10, the epoch by which we wish to see formation of massive black holes. This mild increase in age of this model is compatible with other cosmological considerations such as the shift in the position of the first peak of the CMB power spectrum shown by \cite{elgaroy2005loitering}.\\
The other cosmological models are sufficiently close to $\Lambda$CDM model in the context of age. The DDE model constrained by PDDS increases the age by 4$\%$ of $\Lambda$CDM near present epoch. At high redshift this increment reduces to about 0.3$\%$. This is smaller than the age increment in braneworld cosmology. The $\omega$CDM cosmology, similarly produces only mild increment in age. We consider these alternatives, therefore, as viable cosmologies for examining the time scales of massive black hole formation and seed masses required for these black holes.\\

\section{Mass of seeds in various cosmologies}

We consider both Eddington limited and super Eddington accretion onto seed black holes. The accretion rate and seed mass are expressed in equation 3 and equation 4 respectively. The Eddington fraction $f$ is governed by the black hole spin $\chi$ and the accretion ratio. For standard Eddington limited accretion (where accretion ratio $a$ = 1), $f$ = 1. For accretion to occur at super-Eddington rate with varying black hole spin ($\chi$), $a$ and $f$ has to be greater than 1. Using equation (5) for different choices of the accretion ratio, $f$ is calculated with varying black hole spin.\\
For spinning black holes with different choices of accretion ratios, the Eddington fraction approximately converges to 2.4 as black hole spin $\chi$ tends to the Kerr limit $\chi$ = 1 (see Figure 4). \cite{pezzulli2016super} shows that in a model with black hole feedback, a high accretion rate is maintained at z $\geq$ 10. Also, the  gas inflows from larger scales, where gas is not constrained by the black hole's gravity, may be suppressed by radiation heating. The average rate of the accretion ratio is restricted to about 10, due to the radiation heating effect from a black hole, which causes the accretion behavior to be episodic \citep{ciotti2001cooling}. Hence in the case of super Eddington accretion($f>1$), our choices for accretion ratio are limited to 2 and 5, where the Eddington fraction is just greater than 1, for two values of black hole spin $\chi$ = 0.5 and 0.9.\\
Our choices for the final black hole masses $M_{BH}(t(z_{f}))$ (in equation 4) are $10^8 M_{\odot}$ and $10^9 M_{\odot}$ which are considered to be formed by redshift z = 10. The seed masses required to give rise to SMBHs are calculated for different values of radiative efficiencies $\epsilon$ (0.1 to 0.9) in different cosmologies.\\
To investigate the nature of the black hole seeds, the time slice $\Delta t$ available for the above has been studied in various background cosmologies. The process of growth of such massive black holes has to be initiated well before the epoch of reionization, possibly within z = 20-30, so as to power quasars at z $>$ 6 \citep{melia2015supermassive}. It is assumed that black hole accretion starts at $z_i$ = 
30. Cosmic age at redshift $z_f$ is expressed as
\begin{equation}
	t(z_f)= \int_{z_{f}}^{\infty} \frac{1}{(1+z)H(z)} 
\end{equation}
The seed masses $M_s$ required for formation of the final black holes $M_{BH}(t(z_{f}))$ with Eddington accretion ($f$ = 1) are displayed in Figure 5. Seed masses for super Eddington limit ($f\ge1$) are calculated and are displayed in Figure 6 and Figure 7. For the both the black hole spin values ($\chi$=0.5 and $\chi$=0.9) the seed masses are calculated for two accretion ratios, $a$ = 2 and 5.\\
In Eddington limited accretion scenario, the seed masses are found to be $\geq 10^{4} M_{\odot}$ for all cosmologies. For super Eddington accretion onto black holes spinning with $\chi$ = 0.5, minimum mass of the seeds are found to be in the range $(10^{2}-10^{3}) M_{\odot}$. And for highly spinning black holes with $\chi$ = 0.9, super Eddington accretion requires seeds of mass in the range $(30-10^{2}) M_{\odot}$. It has been found that, in a particular accretion scenario with fixed radiative efficiency, the seed masses do not strongly depend on cosmologies. Super Eddington accretion onto spinning black holes does reduce the mass of the seeds by a factor $10^{3}$. The fractional difference in seed masses for different cosmologies is displayed in Figure 8 for Eddington limited accretion. We discuss the results on seed masses in section 5.\\
\section{Implication of the massive seed black holes}
The seed masses calculated above interestingly span wide astrophysical window of PBHs. We assume that all these seeds are of primordial origin. If PBHs have a wide mass spectrum they can fulfill three roles as initially proposed by \citep{carr2018primordial}. These are PBHs providing dark matter, binding first clouds and generating the first generation galaxies whose masses are simply related to their central black holes. \\
Here, we study the implication of the massive seeds. \cite{hoyle1966formation} proposed that elliptical galaxies of around $10^{12} M_{\odot}$ can form due to density fluctuation created by mass concentration of $10^{9} M_{\odot}$. It was called a gravitational Coulomb effect. This theory was based on the idea of a steady state cosmology where matter is created by a creation field which is intense near a mass concentration. Matter created thus expand outward to manifest as cosmic expansion. However, if there is sufficient mass concentration at the center causing mass fluctuation, then expansion can turn around and form a bound system. In current context, a massive seed black hole ($\sim 10^{7}-10^{9}) M_{\odot}$ can assemble a dark matter halo around it which further accretes gases to form galaxies \citep{carr2018primordial, liu2022accelerating}. This is called "seed effect". Galaxies can also form via Poisson density fluctuation which gets added to the primordial adiabatic fluctuations \citep{liu2022accelerating, carr2018primordial}.\\
\cite{labbe2023population} have derived the cumulative stellar mass density at z = 8 and z = 10 using the 14 galaxy candidates with masses around $\sim 10^{9}-10^{11} M_{\odot}$ found by JWST. \cite{liu2022accelerating} have considered an analytical model based on linear perturbation theory and Press Schecther formalism and evaluated the parameter space for the product $m_{PBH}f_{PBH}$, with the help of 'Poisson effect' and 'seed effect'. If the matter power spectrum P(k) is known, the classical Press- Schechter formalism \citep{press1974formation} can be used to calculate halo mass function dn(M,z)/dM. This allows one to calculate stellar mass density (commoving) for galaxies above certain stellar mass $M_{*}$ as\\
\begin{equation}
    \rho_{*}(>M_{*},z) = \epsilon_{*} f_{b} \rho_{h} (>M_{h},z)
\end{equation}
where $\epsilon_{*}=M_{*}/f_{b}M_{h}$ is the star formation efficiency (SFE), $f_{b}$ is the cosmic baryon fraction, $M_{h}$ is the halo mass and $\rho_{h}$ is the halo mass density,
\begin{equation}
    \rho_{h} = \int_{0}^{M}{M\left(\frac{dn(M,z)}{dM}\right)}dM
\end{equation}
The Poisson effect produces isocurvature fluctuations whose power spectrum depends on PBH mass ($m_{PBH}$) and PBH dark matter fraction ($f_{PBH}$). \cite{liu2022accelerating} found the criteria on the parameter $m_{PBH}f{PBH}$ which satisfies existence of massive galaxies at very high redshift. Low and high SFE ($\epsilon_{*}$ = 0.1 and 1) were chosen. For Poisson effect, the bound on $m_{PBH}f_{PBH}$ is found as 
\begin{equation}
\begin{aligned}
    m_{PBH}f_{PBH} \geq 6 \times10^{6} M_{\odot},  \epsilon_{*}=0.1\\
    m_{PBH}f_{PBH} \geq 2 \times10^{5} M_{\odot},  \epsilon_{*}=1
\end{aligned}
\end{equation}
For seed effect the above bounds are as follows
\begin{equation}
\begin{aligned}
    m_{PBH}f_{PBH} \geq 3 \times10^{5} M_{\odot}, \epsilon_{*}=0.1\\
    m_{PBH}f_{PBH} \geq 3 \times10^{3} M_{\odot}, \epsilon_{*}=1
    \end{aligned}
\end{equation}
Number density of PBHs is given by \citep{liu2022accelerating}
\begin{equation}
    n_{PBH}= \left(\frac{f_{PBH}}{m_{PBH}}\right)(\Omega_{m}-\Omega_{B})\rho_{crit}
\end{equation}
where $\rho_{crit}=3H^{2}_{0}/8\pi G$ is the critical density. We consider the $\Lambda$CDM parameters \citep{aghanim2020planck}, $\Omega_{m}=0.31, \Omega_{b}=0.0493$ and $H_{0}$ = 67.36 $kms^{-1}Mpc^{-1}$.\\
If the results of \cite{labbe2023population} are an overestimate of the stellar mass density at z $\sim$ 10, then smaller values of SFE ($\epsilon_{*}$) in different PBH models can help to explain the recent observations. For a reduced stellar masses by 1.6 dex, \cite{steinhardt2023templates} has shown that for the Poisson effect $m_{PBH}f_{PBH} \geq 2\times 10^{5} M_{\odot}$ for $\epsilon_{*} \leq$ 0.025 and for seed effect $m_{PBH}f_{PBH} \geq 200 M_{\odot}$ for $\epsilon_{*}$ = 0.1 and $m_{PBH}f_{PBH} \geq 2 M_{\odot}$ for $\epsilon_{*}$ =1.\\
We consider the seed mass $(10^{5}-10^{8}) M_{\odot}$ , generated in this study as massive PBHs. For different choices of SFE ($\epsilon_{*}$) and the given bounds on $m_{PBH}f_{PBH}$, the PBH dark matter fraction and their number densities are calculated in both Poisson effect and seed effect and displayed in Figure 8. It is found that, for the Poisson effect, the number densities and correspondingly the $f_{PBH}$ values for the PBHs of the above mass in the range ($10^{5}-10^{8}$) $M_{\odot}$ are quite high (see Figure 9(a)). However, in the seed effect $n_{PBH}$ and $f_{PBH}$ decrease for massive PBHs (see Figure 9(b)). The lower value of $m_{PBH}f_{PBH}$ is related to reduced stellar mass, for which the number density of massive PBHs decreases. It is observed that in seed effect, massive PBHs with higher value of SFE and reduced stellar mass (relative to \cite{labbe2023population}) can yield number densities, $n_{PBH} \leq 10^{-2} Mpc^{-3}$.  

\subsection{PBHs seeded halos and black hole to stellar mass ratio}
Here we explore the evolution of the cosmic structures with the help of PBHs as seeds. The idea is that extremely massive PBHs can act as seeds for the high-z SMBHs that leads to evolution of massive galaxies through the seed effect or Poisson effect. We consider the seed effect. To obtain gas and stellar mass around the seed PBH, we adopt the method of \cite{dayal2024exploring}.\\
In seed effect a PBH assembles a dark matter halo of mass $M_{h}$. The average halo accretion rate as a function of redshift is given by \citep{trac2015scorch}
\begin{equation}
    <\dot{M_{h}}> = 0.21\left(\frac{M_{h}}{10^{8}M_{\odot}}\right)^{1.06}\left(\frac{1+z}{7}\right)^{2.5}\left[\frac{M_{\odot}}{yr}\right]
\end{equation}
For halo formation in spherical top-hat collapse model, the virial temperatures of bound halos is given by \citep{loeb2001reionization}
\begin{equation}
    T_{vir} = 1.98 \times 10^{4} \left(\frac{\mu}{0.6}\right)\left(\frac{M_{h}}{10^{8}h^{-1}M_{\odot}}\right)^{2/3}\left(\frac{\Omega_{m}\Delta_{c}}{\Omega^{z}_{m}18\pi^{2}}\right)\left(\frac{1+z}{10}\right)K
\end{equation}
Here $\Delta_{c}$ is the overdensity relative to the critical density at the epoch of collapse. It is given by \citep{loeb2001reionization}
\begin{equation}
   \Delta_{c}=18\pi^{2}+82d-39d^{2} 
\end{equation} 
The function d is expressed as d = $\Omega^{z}_{m}-1$ which is evaluated at collapse epoch z with 
\begin{equation}
    \Omega^{z}_{m}=\frac{\Omega_{m(z)}(1+z)^{3}}{{E^{2}(z)}}
\end{equation}
Here $E^{2}(z)= H^{2}(z)/H^{2}_{0}$. Our cosmological models used for calculation of seed masses enter into $\Omega^{z}_{m}$ through the H(z) function. In classical Einstein-de-Sitter cosmology, $E^{2}(z)=\Omega(1+z)^{3}$ with $\Omega_{m}=1$. This gives d=0 and overdensity $\Delta_{c}=18\pi ^{2}$.\\
When cold dark matter particles drive the gravitational potential of a virialized object, the over-density of baryons can be expressed as \citep{dayal2024exploring}\\
\begin{equation}
    \delta_{b} =\left(1+ \frac{6T_{vir}}{5\Bar{T}}\right)^{3/2},
\end{equation}
where $\Bar{T}$ is the background temperature given as \citep{loeb2001reionization}, $\Bar{T}=170\left[\frac{1+z}{100}\right]^{2}$K. Once the halo starts assembling and exceeds a minimum mass $M^{min}_{h}$, it starts accumulating baryonic matter maintaining the cosmological baryon fraction $f_{b}$. We express evolution of gas content inside a dark matter halo with redshift as
\begin{equation}
\frac{dM_{g}}{dz} = f_{b}\frac{dM_{h}}{dz},    
\end{equation}
The total mass accumulated within the PBH seeded dark matter halo between epochs $z_{1}$ and $z_{2}$ ($z_{2}<z_{1}$) is
\begin{equation}
    dM_{g} = f_{b}\int_{z_{1}}^{z_{2}}\left(\frac{d M_{h}}{dz}\right) dz
\end{equation}
The halo accretion rate in equation (29) can be written in terms of redshift as
\begin{equation}
    \frac{dM_{h}}{dz}=0.21\left(\frac{M_{h}}{10^{8}M_{\odot}}\right)^{1.06}\left(\frac{1+z}{7}\right)^{{2.5}}\left(\frac{M_{\odot}}{yr}\right)\left(\frac{dt}{dz}\right)
\end{equation}
The time gap between epochs $z_{1}$ and $z_{2}$ is given by the cosmology dependent relation
\begin{equation}
    dt/dz = \frac{-1}{H(z)(1+z)}
\end{equation}
Taking $z_{1}$ = z and $z_{2}$ = 10, we find the following expression for the total gas mass accumulated within the halo.
\begin{equation}
    \Delta M_{g}\approx1.62\times10^{7}f_{b}h^{-1}M_{\odot}\int_{10}^{z}(1+z)^{3/2}\left(\frac{M_{h}}{10^{8}M_{\odot}}\right)^{1.06}\frac{dz}{E(z)} 
\end{equation}
Here $h=H_{0}/100 kms^{-1}Mpc^{-1}$. Using equation (25) in equation (28) we determine the minimum halo mass $M^{min}_{h}$ that can host gas mass with a baryonic overdensity, $\delta_{b}=200$. We adopt $\mu \approx 1.22$ for primordial neutral gas. Using $M^{min}_{h}$ in equation (33) and taking cosmic baryon fraction as $f_{b}\approx0.15$ we calculate $\Delta M_{g}$ for different cosmological models and display the variation of $\Delta M_{g}$ with redshift from $z_{2}=10$ to $z_{1}=30$ in Figure 10.\\
The central PBH inside the dark matter halo accretes a fraction of this gas content at Eddington or super-Eddington rate to grow into a SMBH. A fraction of the gas mass is now allowed to form stars with efficiency $\epsilon_{*}$ ranging from 0.1 to 1. We take the initial epochs as z = 30. The stellar mass in a galaxy with gas mass $\Delta M_{g}$ is given by
\begin{equation}
    M_{*} = \epsilon \Delta M_{g}
\end{equation}
The ratio of black hole to stellar mass ($M_{BH}/M_{*}$) is calculated for the final black hole masses $M_{BH} = 10^{8} M_{\odot}$ and $10^{9} M_{\odot}$ and is shown in Figure 11. We compare the calculated ratios with the values reported by various observations displayed in table 1.\\

\section{Results and Discussion}

The nature of seeds for SMBHs with masses $10^{8}$ and $10^{9} M_{\odot}$ in various background cosmologies has been studied in this work. The growth of SMBHs by $z_{f}=10$ is studied through Eddington and super Eddington accretion onto seeds present at a primordial epoch ($z_{i}=30$). For super Eddington accretion, we considered rotating seed black hole with spin $\chi$= 0.5 and 0.9, with two different choices of accretion ratio $a$ = 2 and 5. The choice of radiative efficiency $\epsilon$ is made as (0.1-0.9) for both Eddington and super Eddington accretion. The growth of $(10^{8}-10^{9}) M_{\odot}$ black holes is studied in three GR based cosmologies: $\Lambda$CDM, $\omega$CDM, DDE with CPL parametrization for two datasets: SNeIa only and Planck+DESI+DESY5 (PDDS) and one non GR cosmology namely the braneworld model. In the following discussion we discuss the seed mass range as (X-Y) where X and Y represent seed masses respectively for final black hole mass $10^{8} M_{\odot}$ and $10^{9} M_{\odot}$. \\
In case of $\Lambda$CDM the minimum mass of seed black holes required to form SMBHs via Eddington accretion is $M_{s}= 5\times10^{4} M_{\odot}$ and upwards (see Figure 5). Whereas, for super Eddington accretion onto a rotating seed black hole of spin $\chi=0.5$, the mass of seeds is within $(10^{4}-10^{5}) M_{\odot}$ for accretion ratio $a$=2 and  $(900-10^{3}) M_{\odot}$ for $a$=5 (see Figure 6). For $\chi=0.9$, seed masses for $a=2$ are  $(10^{3}-10^{4}) M_{\odot}$ and those for $a$=5 reduce to $(60-600) M_{\odot}$ (see Figure 7 for reduction of seed masses for $a=5$).\\
In $\omega$CDM cosmology with EoS $\omega$ = -0.80, the seeds required to grow within Eddington limited accretion are found to be in the range 4$\times 10^{4} M_{\odot}$ and upwards (see Figure 5). In case of super Eddington accretion, the minimum mass of seeds with $\chi$=0.5 are found to be in the range $(10^{4}-10^{5}) M_{\odot}$ for $a$=2 and  $(600-10^{3}) M_{\odot}$ for $a$=5 (see Figure 6).For seeds with $\chi$=0.9, the masses decrease to $(10^{3}-10^{4}) M_{\odot}$ for $a$=2 and $(30-400) M_{\odot}$ for $a$=5 (see Figure 7).\\
In DDE with CPL parametrization, the masses of seeds within Eddington limited accretion are found to be $6\times10^{4} M_{\odot}$ and upwards in PDDS dataset and $3\times10^{5} M_{\odot}$ and upwards in SNeIa only dataset (see Figure 5). In the SNeIa only dataset, the seed mass range for $\chi$=0.5 is found to be $(10^{5}-10^{6}) M_{\odot}$ for $a$=2 and $(10^{4}-10^{5}) M_{\odot}$ for $a$=5  (see Figure 6). The black hole seed masses for $\chi$=0.9 are found to be in the range $(10^{4}-10^{5}) M_{\odot}$ for $a$=2 and $(900-10^{4}) M_{\odot}$ for $a$=5 (see Figure 7). But for the PDDS dataset, the seed masses required to grow into $10^{8}-10^{9} M_{\odot}$ black holes in both the accretion scenario are found to be identical to those of $\Lambda$CDM. This parallelism between the two models is seen for all values of black hole spin and accretion ratio. The reason for similar seed mass range in $\Lambda$CDM and DDE (PDDS) is evident from fractional difference of cosmic timescales in the models which approximates to 0.3$\%$ at high redshift (Figure 3). Thus, DDE (PDDS) model cannot substantially change the black hole seed masses from what is found in $\Lambda$CDM. The DDE(SNeIa only) cosmology causes significant suppression of cosmic age thereby giving extremely massive seed, around $10^{5} M_{\odot}$. A possible explanation for such heavy seeds in this dataset could be the large negative value of EoS. $\omega_{a} = -8.8$ leading to a higher expansion rate both at early and late times.\\
It is worthy of mentioning that the DDE (SNeIa only) cosmology has been improved in the form of DDE (PDDS) having better constraints on the EoS parameters. We keep the earlier version just for illustration of how sensitive the expansion time and hence seed masses can be to the background cosmological parameters. The concluding remarks are based on DDE (PDDS) cosmology only.
In braneworld cosmology, the mass of seeds required to grow into SMBHs within Eddington limit are found to be $> 4\times10^{4} M_{\odot}$ (see Figure 5). In case of growth by super Eddington process, the black hole masses are found to be in the range $(10^{4}-10^{5}) M_{\odot}$ for $a$ = 2 and $(550-10^{3}) M_{\odot}$ for $a$ = 5 for seeds with $\chi=0.5$ (see Figure 6). For $\chi$ = 0.9, the seed masses are drastically reduced. These are in the range $(900-10^{3}) M_{\odot}$ for $a$ = 2 and $(30-300) M_{\odot}$ for $a$ = 5 (see Figure 7).\\
It is realized that growth by accretion at Eddington limit can produce SMBHs with seeds of mass $(10^{4} M_{\odot})$ and upwards for all cosmologies. But with super Eddington accretion the mass of the seed black holes can reduce to few tens of solar mass  to few hundreds of solar mass. It is also observed that a higher black hole spin and a higher accretion ratio is able to generate SMBHs in the range $(10^{8}-10^{9}) M_{\odot}$ with lighter seeds similar to those of LIGO's black holes.\\
We note that although the viable cosmologies considered here can elevate the cosmic time in the early universe, they are unable to strongly distinguish between the seed black hole masses. Many of them produce $\Lambda$CDM like seed masses. Without having cosmologies violating the CMB bound on shift parameter, it is not possible to produce seed masses drastically deviating from the benchmark  $\Lambda$CDM cosmology.\\
Primordial seed black holes within 30 $M_{\odot}$ are believed to form via collapse of high curvature regions in the epoch of meson and baryon formation in the early universe when there was significant reduction of sound speed due to presence of heavy (cold) particles. These black hole masses are potential candidates of PBH dark matter constituting a fraction $f_{PBH}\approx0.003\%$ . Indication of existence of such black holes has come from 5yr OGLE microlensing campaign \citep{niikura2019constraints}. OGLE/GAIA survey also reported a sizable population of black holes in the mass range (1 - 10)$M_{\odot}$ \citep{wyrzykowski2020constraining}. The mass range (10-30)$M_{\odot}$ matches with binary black holes detected by LIGO which opened the new window of stellar mass PBH dark matter (\cite{abbott2016binary}; \cite{abbott2019binary}; \cite{bird2016did}; \cite{clesse2017clustering}). Interestingly baryon accretion onto stellar mass seed black holes with mass (1-100)$M_{\odot}$ is now believed as potential source of Cosmic X-ray Background (CXB) and Cosmic Infrared Background (CIB) (\cite{hasinger2020illuminating}; \cite{ziparo2022cosmic}).\\
We feel it necessary to put a remark on low mass seeds. The role of black hole spin on seed mass needed for growth onto massive black holes by z=10 carries prospect for future gravitational wave astronomy and PBH mass accretion. There are predictions that PBHs with mass $>O(30 M{_\odot})$ can have very large spin \citep{de2020evolution}. In this case growth of our seed masses with M$>$ 30 M$_{\odot}$ is an opportunity to study accretion physics in the early universe. Spin and mass have joint contribution to the gravitational waveforms. The above masses, if constitute binaries, are accessible to the ground based gravitational wave detectors such as LIGO, VIRGO and KAGRA.\\
We discuss an interesting implication of higher side of the seed masses, $10^{5}-10^{8} M_{\odot}$. We assume that these seeds are of primordial origin. We consider the constraints on the product $m_{PBH}f_{PBH}$ in both Poisson effect and seed effect, with different values of SFE ($\epsilon_{*}$). We find that in Poisson effect, the number densities ($n_{PBH}$) and corresponding $f_{PBH}$ of ($10^{5}-10^{8}$) $M_{\odot}$ PBHs are extremely large (see figure 9(a)). In seed effect, within the constraints on stellar mass density given by \cite{labbe2023population},($10^{5}-10^{8}$) $M_{\odot}$ PBHs contribute $f_{PBH} \geq 10^{-5}$ with $n_{PBH} \geq 10^{-2} Mpc^{-3}$. If the actual stellar mass density is low relative to the constraints given by \cite{labbe2023population} we find that PBHs with mass above $10^{6} M_{\odot}$ contribute $f_{PBH} \geq 10^{-8}$ and their corresponding number densities are found to be $\geq 10^{-6}$. It is observed that a smaller value of $m_{PBH}f_{PBH}$ is related to reduced stellar mass, for which the number densities also comes down. \\ 
Taking seed effect into account we investigate the growth of baryonic gas content in a dark matter halo seeded by PBH. For baryonic over-density $\delta_{b}=200$ and corresponding minimum halo mass, the evolution of the gas mass ($\Delta M_{g}$) is calculated for all cosmological models and displayed in figure 10. The assembling of the gas inside the dark matter halo is assumed to be initiated at z=30. It is found that the gas content reaches a mass of at least $10^{10} M_{\odot}$ by z=10 in all the cosmologies with an average baryon fraction $f_{b}=0.15$. A fraction of this gas is then allowed to form stars with an efficiency $\epsilon_{*}=0.1$ and 1. The ratio of black hole to stellar mass ($M_{BH}/M_{*}$) is calculated for final black hole masses $10^{8} M_{\odot}$ and $10^{9} M_{\odot}$. For $M_{BH}=10^{9} M_{\odot}$, it is found that $M_{BH}/M_{*}$ = $10^{-1}-10$ for  $\epsilon_{*}$ = 0.1 and $M_{BH}/M_{*}$ = $10^{-2}-1$ for  $\epsilon_{*}$=1 (Figure 11 (a)). Similarly for $M_{BH}=10^{8} M_{\odot}$, it is found that $M_{BH}/M_{*}$ = $10^{-2}-10$ for  $\epsilon_{*}$ = 0.1 and $M_{BH}/M_{*}$ = $10^{-3}-10^{-2}$ for  $\epsilon_{*}$ = 1 (Figure 11 (b)). The range of $M_{BH}/M_{*}$ is obtained for all cosmologies. The derived ranges of black hole to stellar mass ratios are found to be compatible with available observational reports displayed in table 1. \\

\section{Conclusion}
In this work we generate seed black hole masses required to form super massive black holes with mass $10^{8}- 10^{9} M_{\odot}$ by considering cosmic time scales provided by theories of gravity within and outside general relativity, accretion ratio and black hole spin. All the alternatives to $\Lambda$CDM, except the dynamical dark energy model constrained by SNeIa only, produce mild increase in cosmic time relative to the benchmark cosmology which is within tolerance limit demanded by other cosmological constraints. We obtain interesting seed masses for these cosmologies.\\
In super Eddington accretion seed black holes similar to those of LIGO's observations are found for higher seed spin $\chi=0.9$ with high accretion ratio in $\Lambda$CDM, DDE (PDDS) and $\omega$CDM cosmology. Heavy seed black holes are still allowed for low seed spin and low accretion ratio. For Eddington limited accretion, intermediate mass seed black holes (minimum of M$\approx 10^{4} M_{\odot}$) and heavier ones are obtained as common features in all cosmologies.\\
The braneworld cosmology and the DDE (PDDS) cosmology are found to be close to the $\Lambda$CDM cosmology in terms of the seed masses. These cosmologies produce only mild shift in cosmic age at high redshift. We find that these cosmologies are unable to strongly discriminate the seed black hole mass. Many of them are compatible with the seed masses predicted by the standard $\Lambda$CDM cosmology. Some of these seeds are of further interest for gravitational wave astronomy, particularly to be carried by  ground based detectors such as LIGO-Virgo-KAGRA.\\
If massive seeds are of primordial origin we see that seed masses common to the cosmological models and towards the heavier side ($\sim 10^{7}-10^{8}) M_{\odot}$ are eligible to satisfy the criteria of massive galaxy formation through the seed effect provided they contribute small dark matter mass fraction $f_{PBH}$ $\approx 10^{-7}-10^{-3}$. This is however a lower bound of the PBH dark matter fraction. Massive PBHs are likely to contribute a significant fraction of dark matter if one has to address the presence of very high redshift massive galaxies found by JWST. The cosmological models have been able to reproduce black hole to stellar mass ratios which are generally compatible with observational reports.\\
\subsection{Figures and tables}
\begin{table}
    \centering
\begin{tabular}{ccc}
\hline
$M_{\rm BH}/M_\star$ & $z$ & Reference \\
\hline
0.4   & 6.68 & \cite{juodvzbalis2024dormant} \\
0.05--0.1 & 6.25 & \cite{stone2023detection} \\
0.002 & 10.6 & \cite{maiolino2024small} \\
0.01--0.1 & 1--3 & \cite{mezcua2024overmassive} \\
$\sim$0.01 & 4--7 & \cite{pacucci2023jwst} \\
$10^{-3}$--1 & 4--11 & \cite{maiolino2024jades} \\
\hline
\end{tabular}
    \caption{Reported values of $M_{\rm BH}/M_\star$ from observations.}
    \label{tab:placeholder}
\end{table}
\begin{figure}
	\includegraphics[width=\columnwidth]{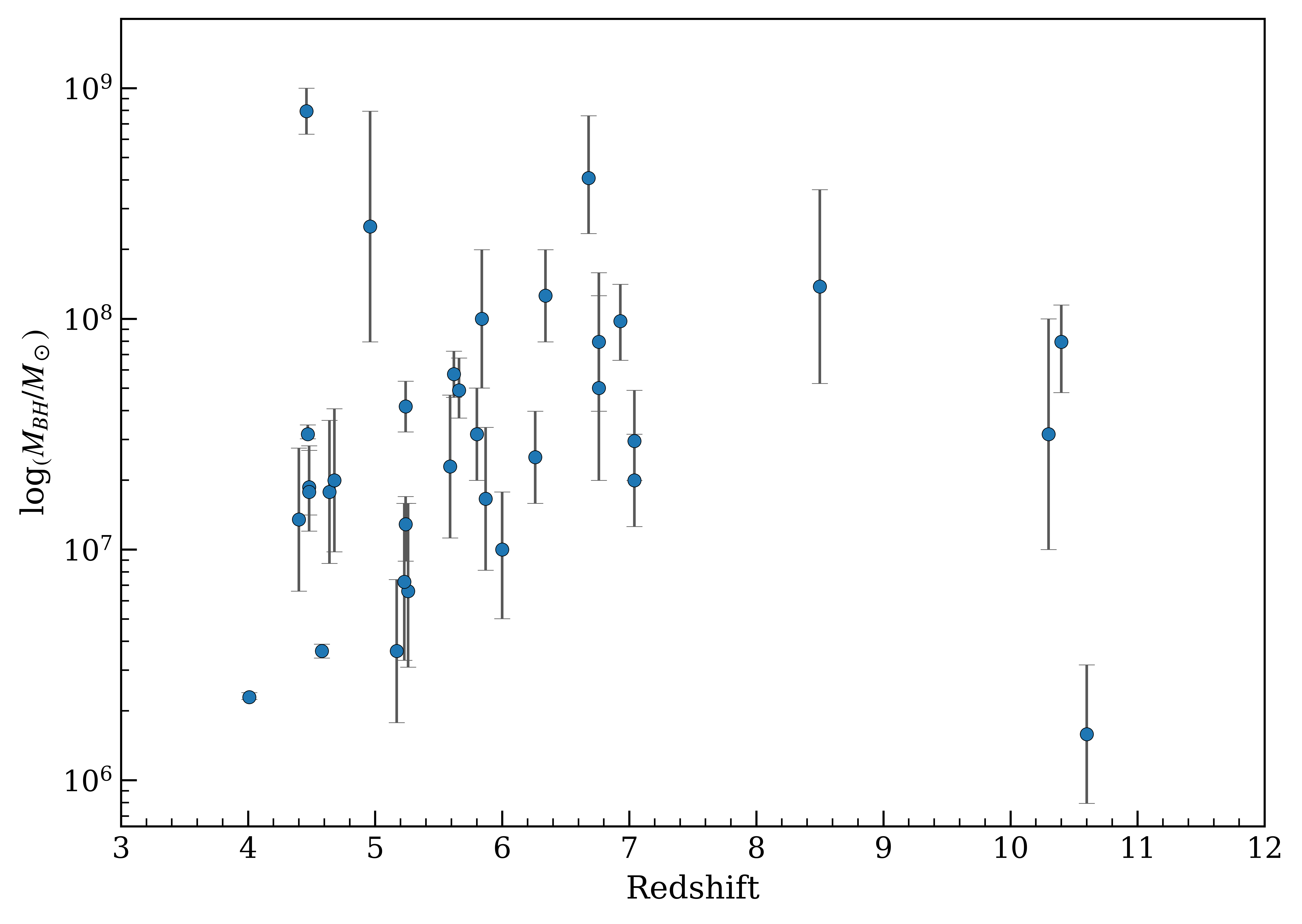}
    \caption{Mass of black hole as a function of redshift for the 34 black holes confirmed by JWST as detailed in Table 1 \citep{dayal2024exploring}. The data points corresponds to asymmetric error bars based on broad Balmer lines, high-ionization lines or X-ray counterparts.}
    \label{fig:example_figure}
\end{figure}
\begin{figure}
	\includegraphics[width=\columnwidth]{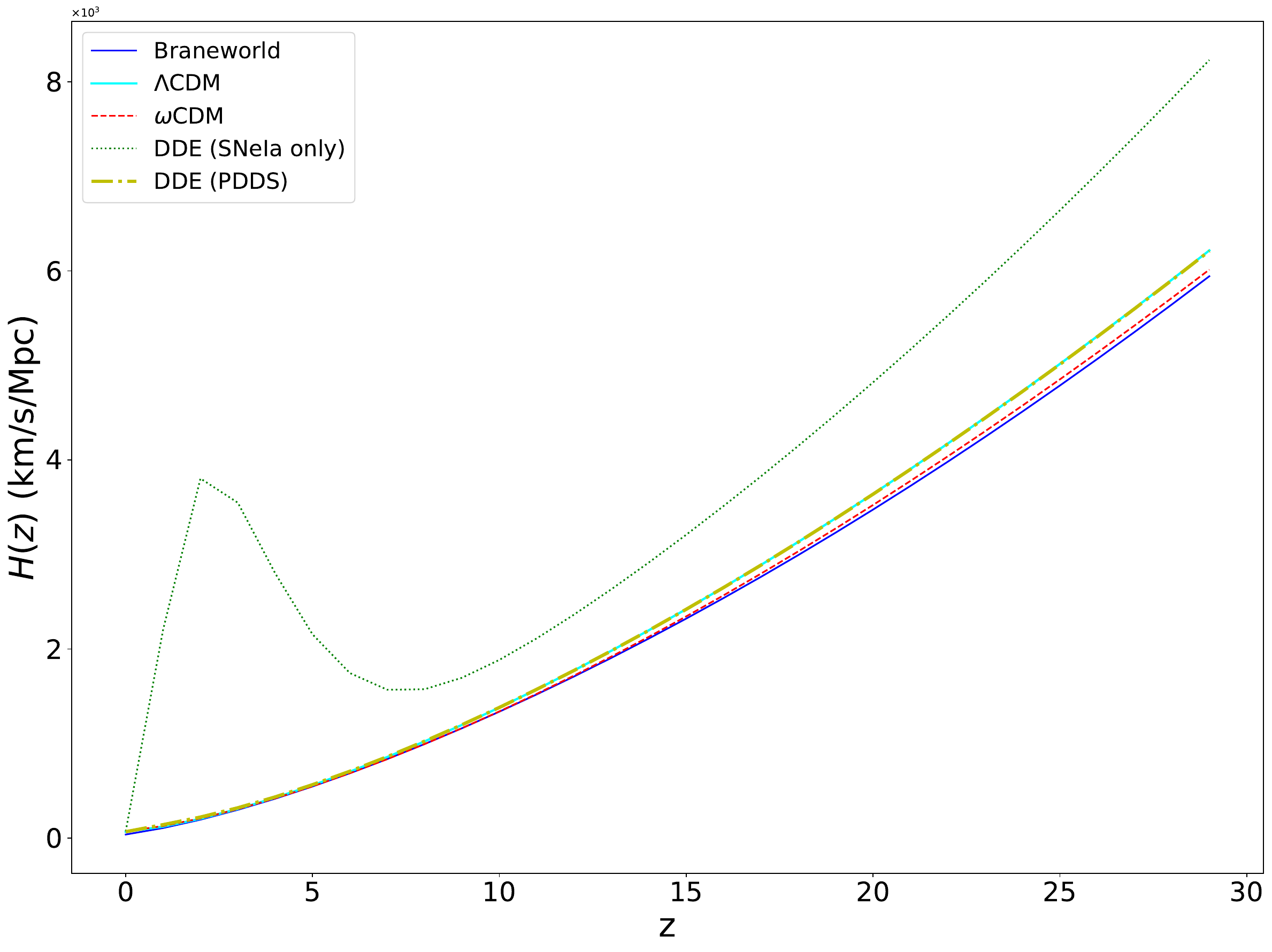}
    \caption{Evolution of the expansion rate H(z) within the redshift slice (0-30) in various cosmologies}
    \label{fig:example_figure}
\end{figure}
\begin{figure}
	\includegraphics[width=\columnwidth]{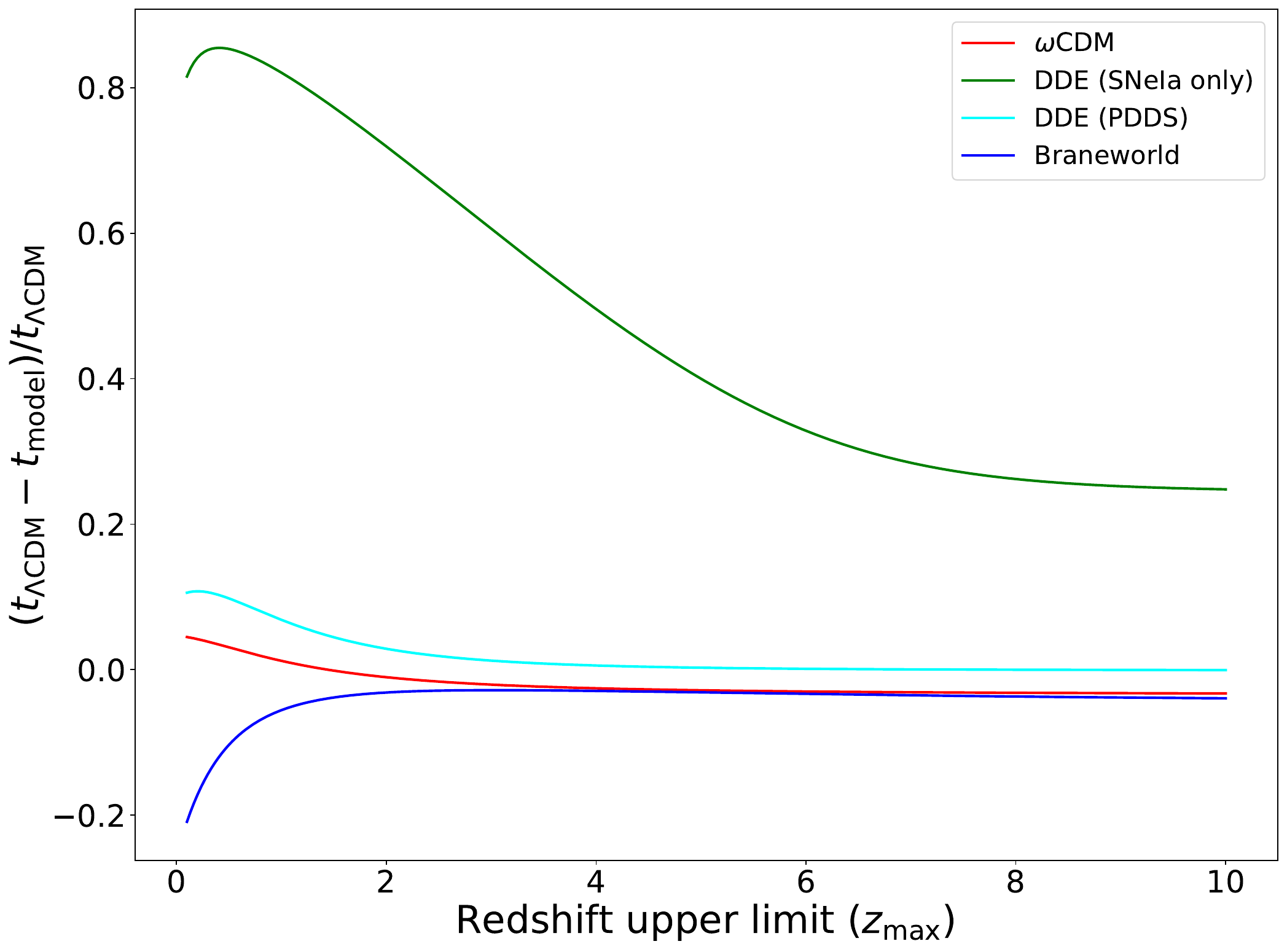}
    \caption{Fractional difference of cosmic age relative to the $\Lambda$CDM}
    \label{fig:example_figure}
\end{figure}
\begin{figure}
	\includegraphics[width=\columnwidth]{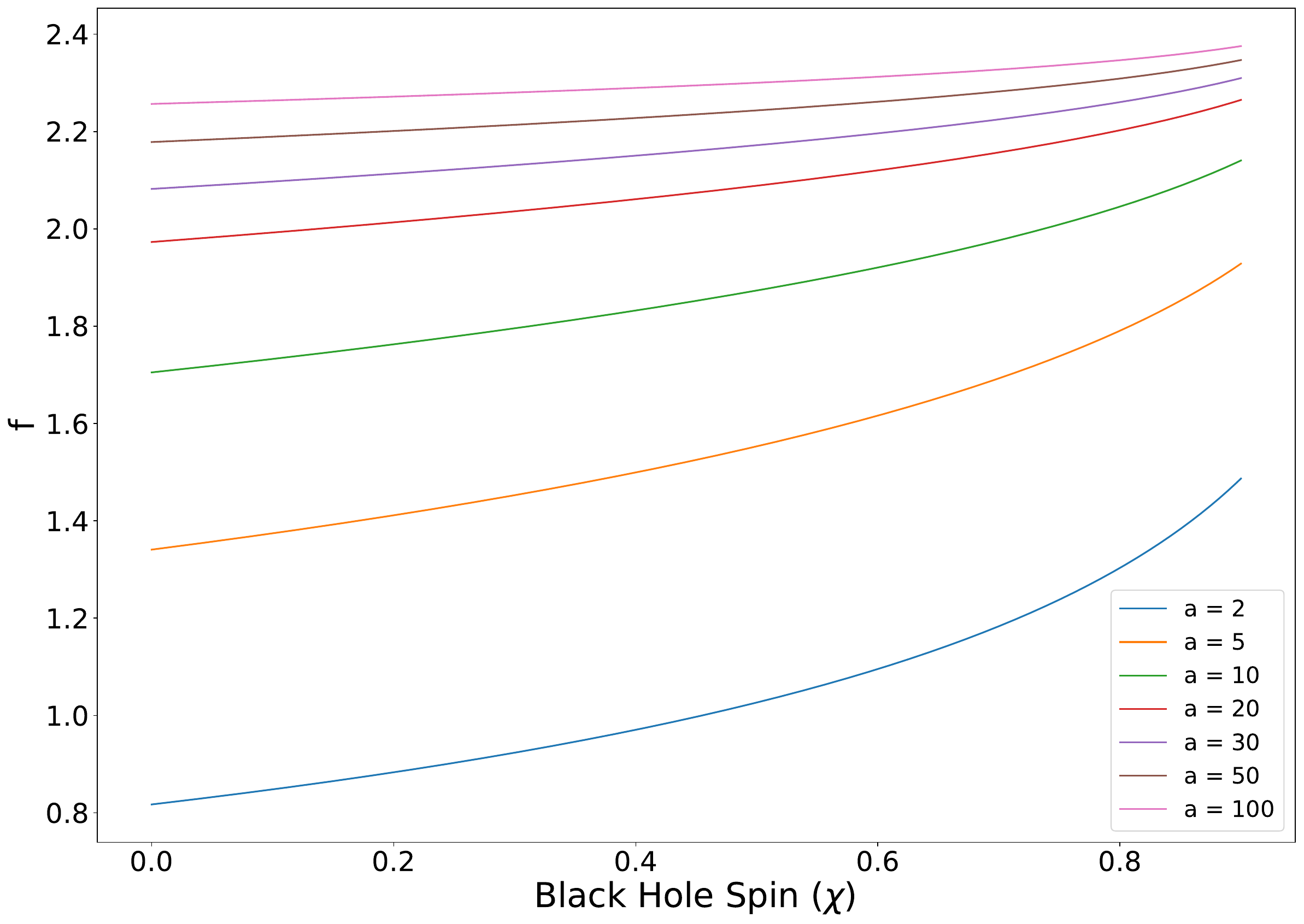}
    \caption{Eddington fraction ($f$) vs black hole spin $(\chi)$ for different choices of the accretion ratio}
    \label{fig:example_figure}
\end{figure}
\begin{figure}
	\includegraphics[width=\columnwidth]{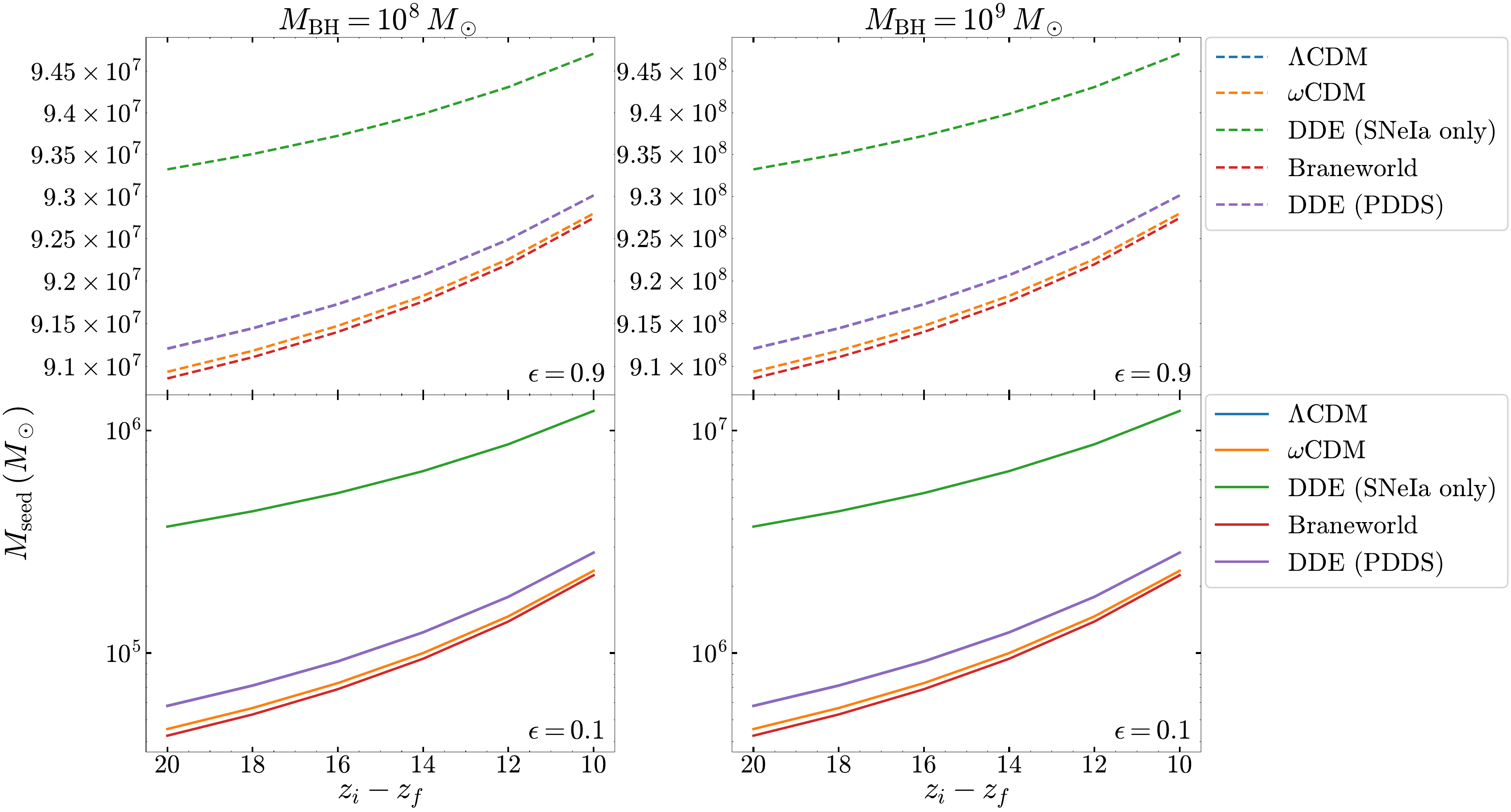}
    \caption{Seed masses required to form $(10^{8}-10^{9}) M_{\odot}$ black holes within the redshift slice $(z_{i}-z_{f})$ allowed for growth within Eddington limit in various cosmologies.}
    \label{fig:example_figure}
\end{figure}
\begin{figure}
	\centering
	\begin{subfigure}[b]{0.50\textwidth}
		\centering
		\includegraphics[width=\textwidth]{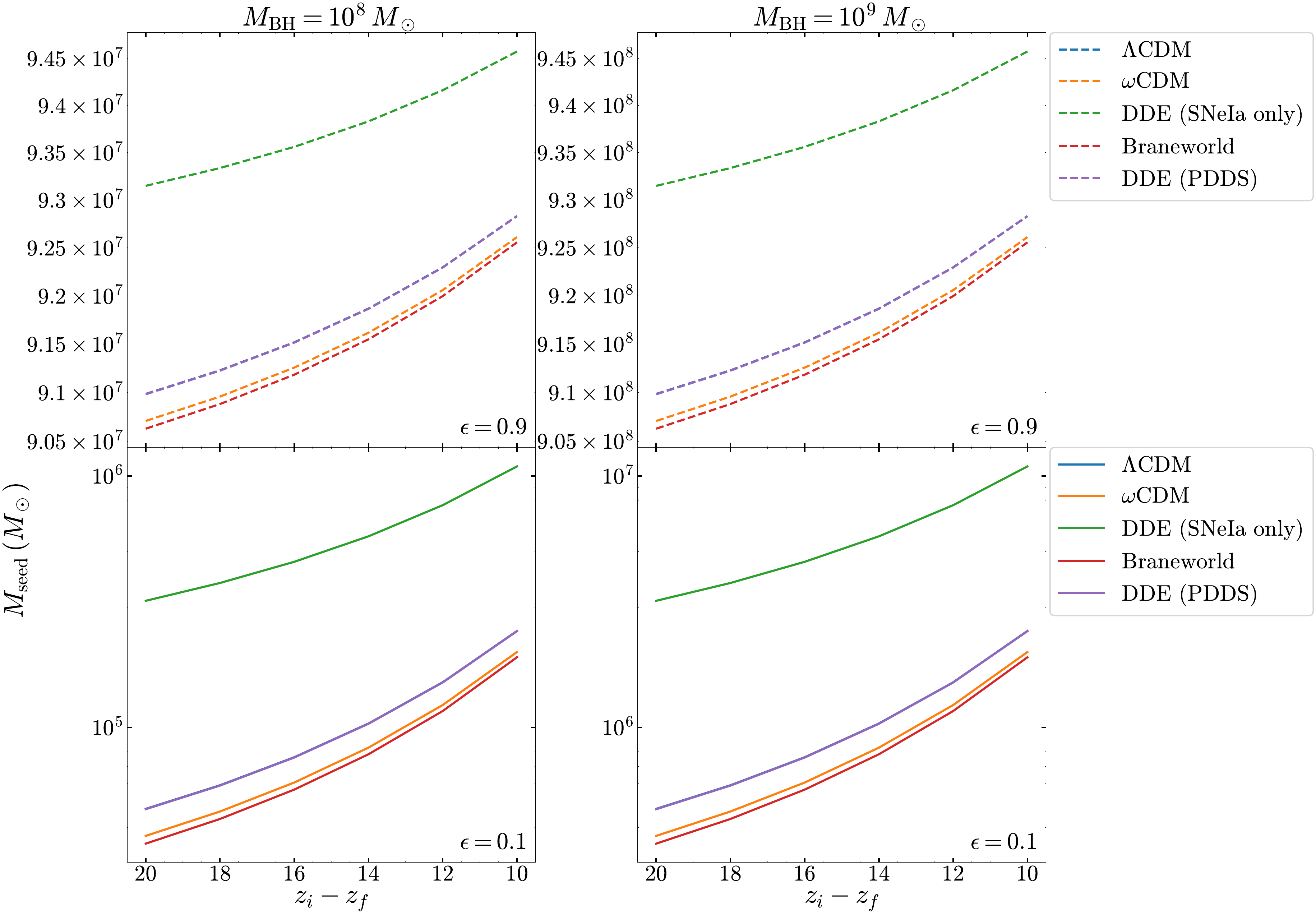}
		\caption{}
	\end{subfigure}
	\\
	\begin{subfigure}[b]{0.50\textwidth}
		\centering
		\includegraphics[width=\textwidth]{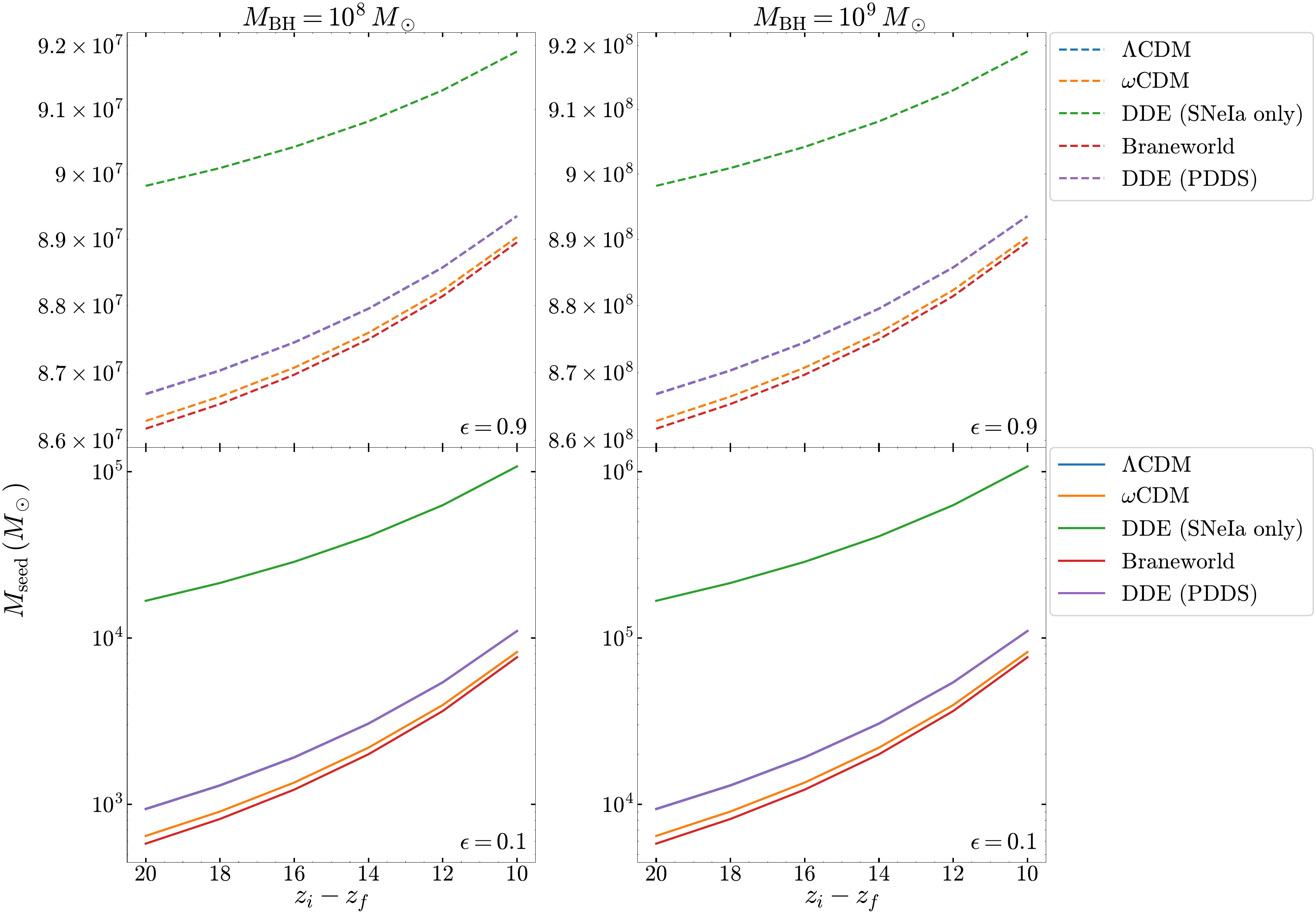}
		\caption{}
	\end{subfigure}
	\caption{Seed masses required to form $(10^{8}-10^{9}) M_{\odot}$ black holes within the redshift slice $(z_{i}-z_{f})$ allowed for super Eddington growth with $\chi =0.5$ (a) accretion ratio=2, (b) accretion ratio=5 in various cosmologies.}
\end{figure}   

\begin{figure}
	\centering
	\begin{subfigure}[b]{0.50\textwidth}
		\centering
		\includegraphics[width=\textwidth]{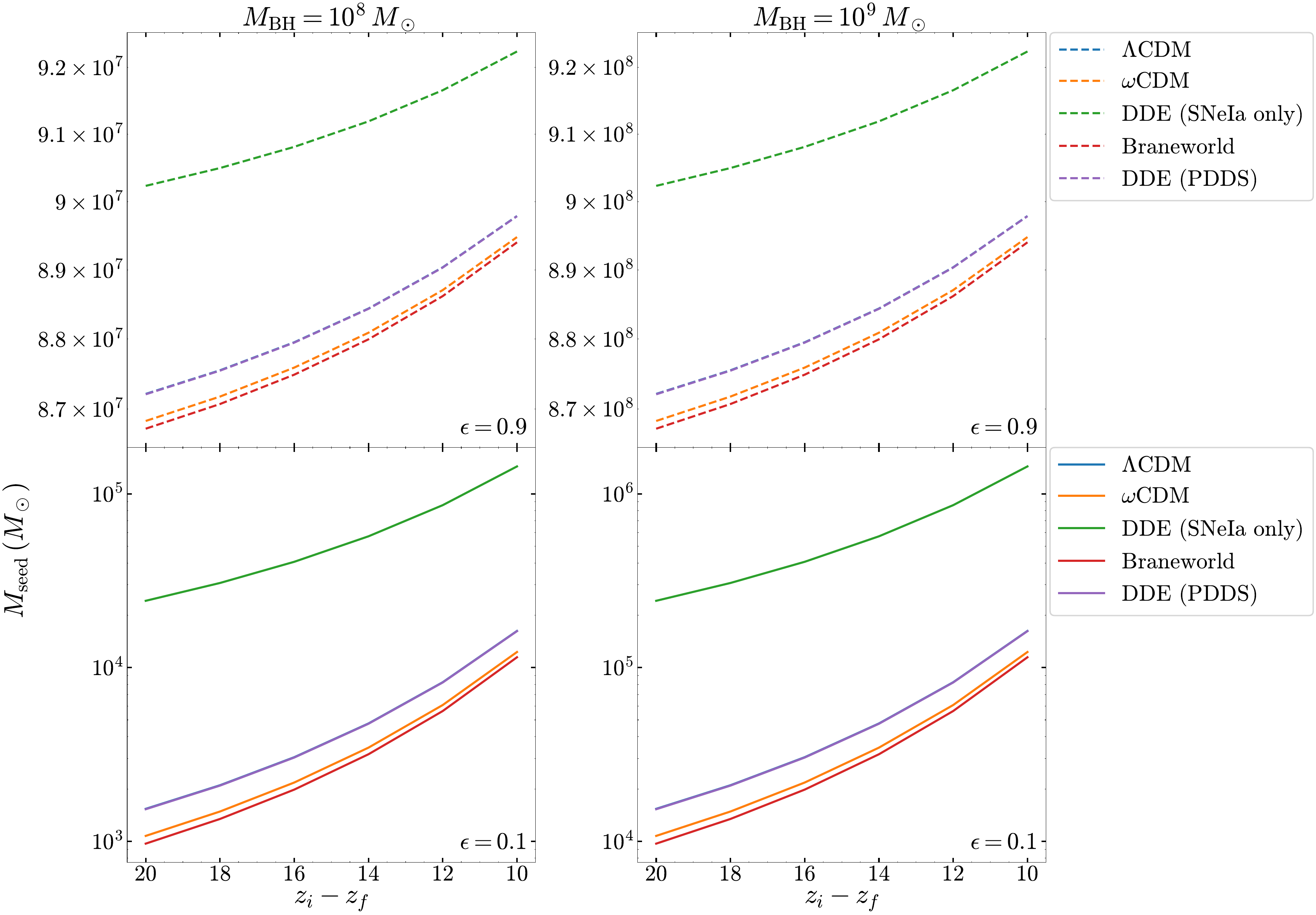}
		\caption{}
	\end{subfigure}
	\\
	\begin{subfigure}[b]{0.50\textwidth}
		\centering
		\includegraphics[width=\textwidth]{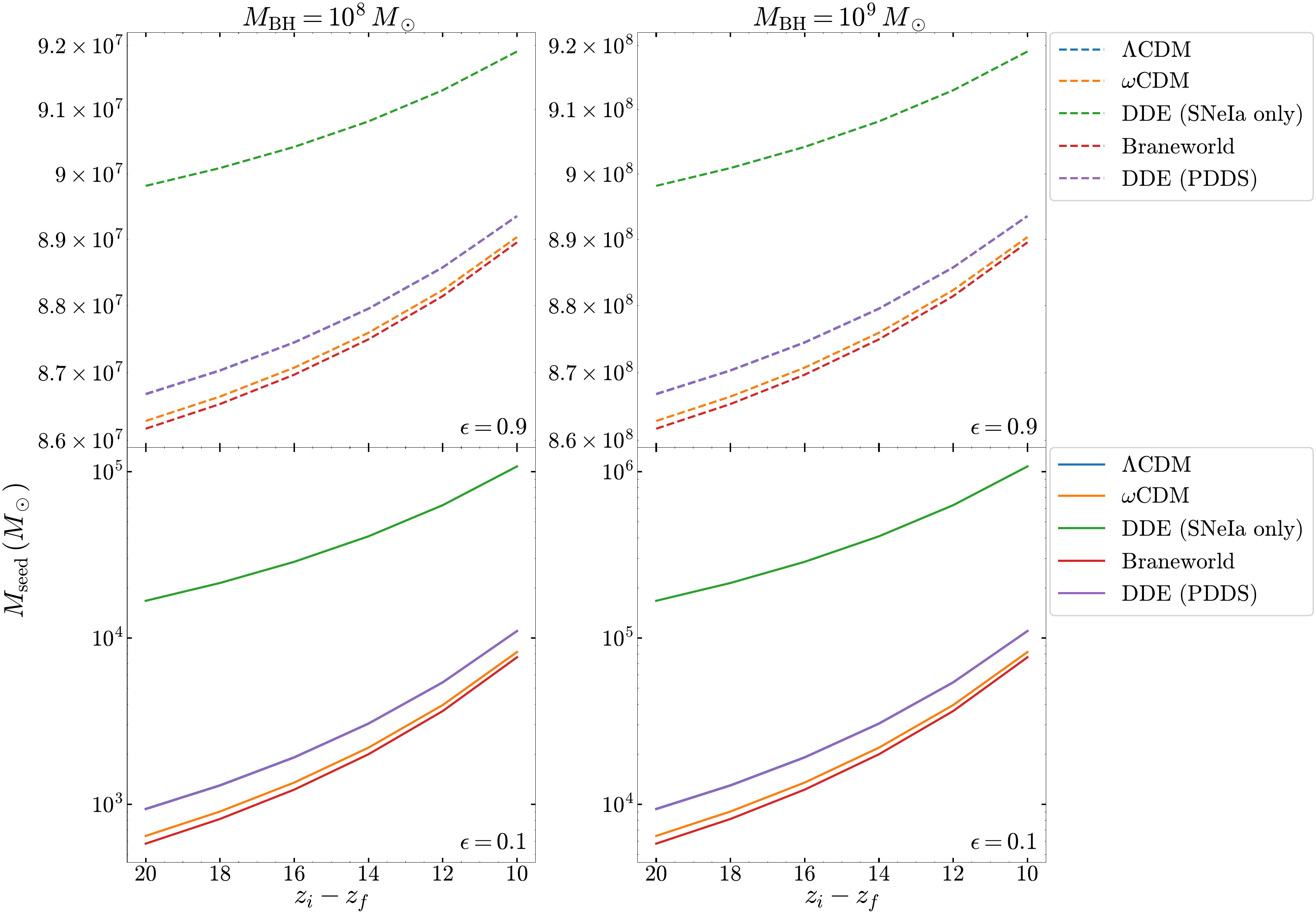}
		\caption{}
	\end{subfigure}
	\caption{Seed masses required to form $(10^{8}-10^{9}) M_{\odot}$ black holes within the redshift slice $(z_{i}-z_{f})$ allowed for super Eddington growth with $\chi =0.9$ (a) accretion ratio=2, (b) accretion ratio=5 in various cosmologies.}
\end{figure}
\begin{figure}
	\includegraphics[width=\columnwidth]{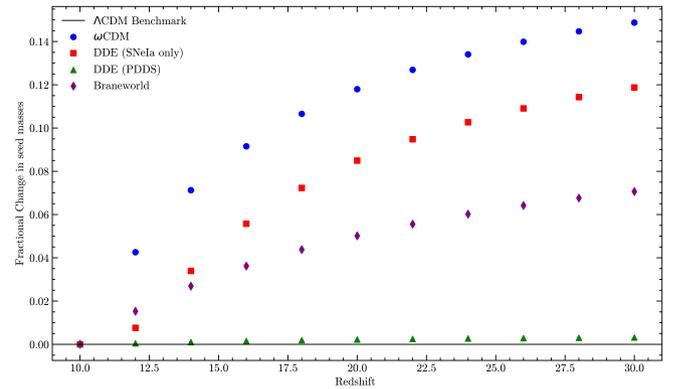}
    \caption{Difference in seed masses calculated in various cosmologies and seed masses in $\Lambda$CDM within the redshift slice $(z_{i}-z_{f})$ allowed for growth to $(10^{8}-10^{9}) M_{\odot}$ black holes within Eddington limit.}
    \label{fig:example_figure}
\end{figure}
\begin{figure}
	\centering
	\begin{subfigure}[b]{0.45\textwidth}
		\centering
		\includegraphics[width=\textwidth]{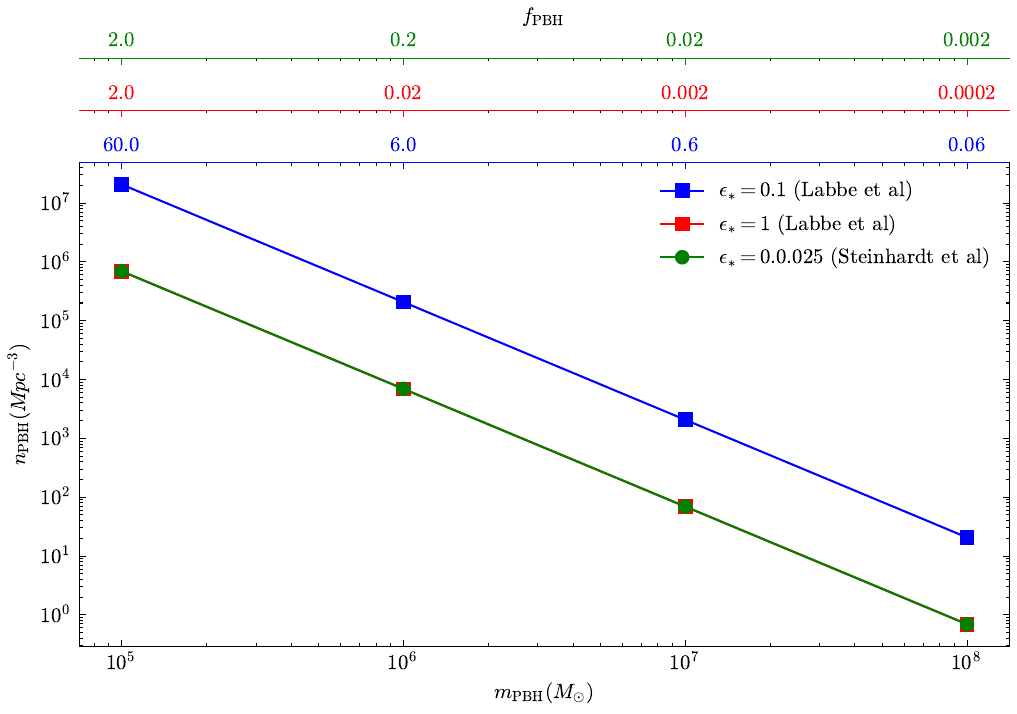}
		\caption{}
	\end{subfigure}
	\\
	\begin{subfigure}[b]{0.45\textwidth}
		\centering
		\includegraphics[width=\textwidth]{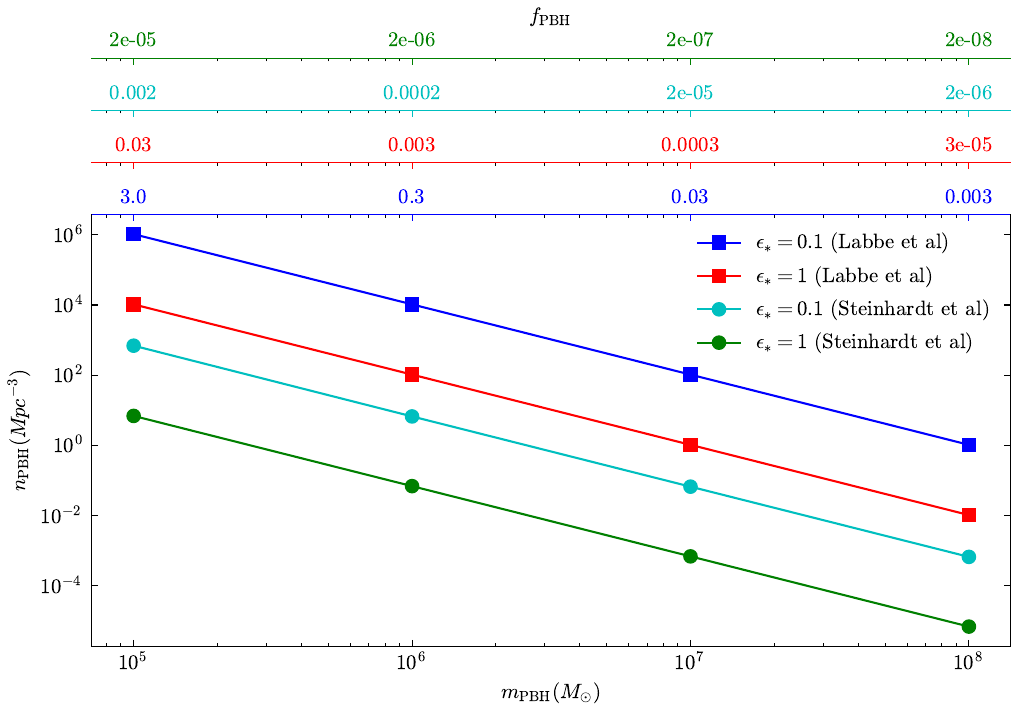}
		\caption{}
	\end{subfigure}
	\caption{Number densities of PBHs ($n_{PBH}$) corresponding to $f_{PBH}$ for the mass range $(10^{5}-10^{8})M_{\odot}$ in the (a) Poisson effect and (b) seed effect. The top axes displays the various $f_{PBH}$ values, corresponding to the PBH mass range, calculated from constraints derived from stellar density results of \citet{labbe2023population} and \citet{steinhardt2023templates}}
\end{figure}
\begin{figure}
    \centering
    \includegraphics[width=\columnwidth]{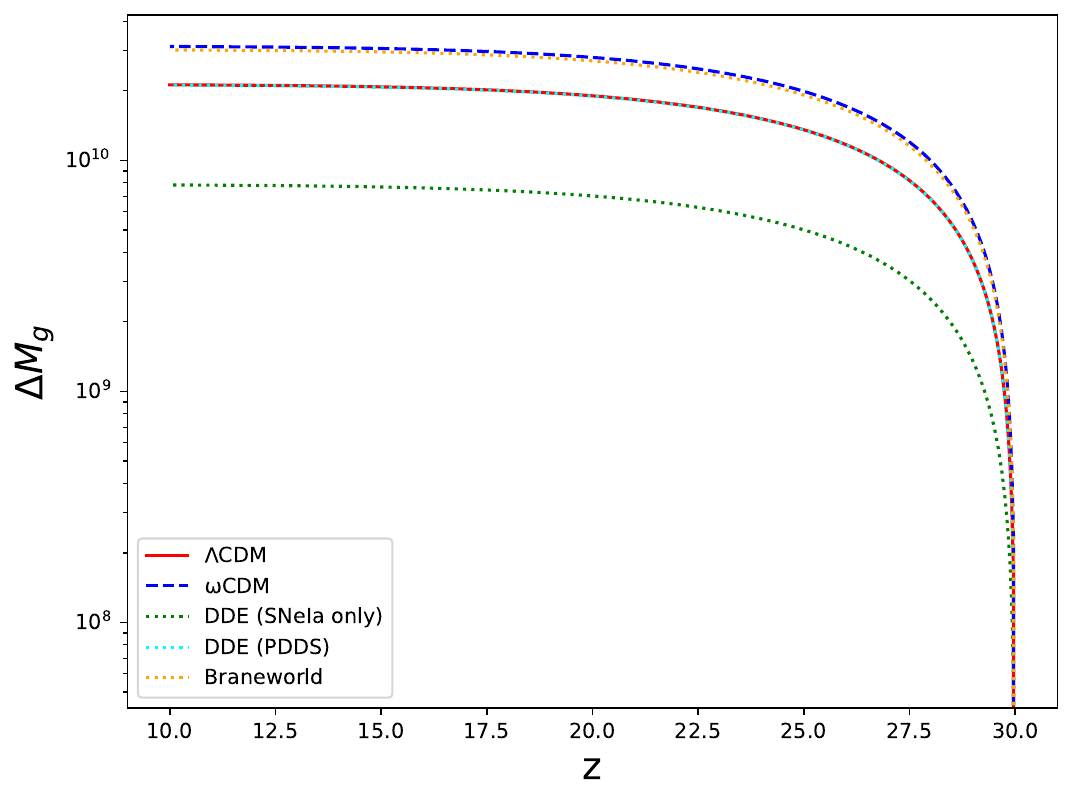}
    \caption{Evolution of the accreted gas mass $M_{g}$ with redshift inside a PBH seeded dark matter halo.}
    \label{fig:placeholder}
\end{figure}
\begin{figure}
	\centering
	\begin{subfigure}[b]{0.45\textwidth}
		\centering
		\includegraphics[width=\textwidth]{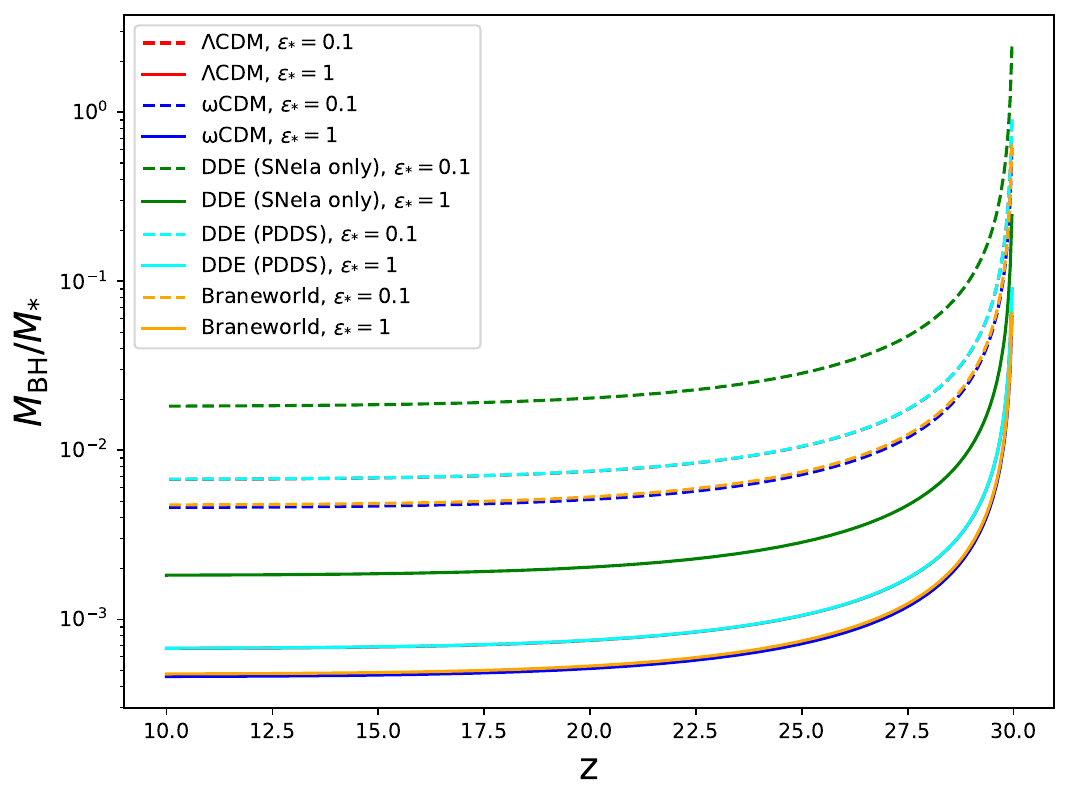}
		\caption{}
	\end{subfigure}
	\\
	\begin{subfigure}[b]{0.45\textwidth}
		\centering
		\includegraphics[width=\textwidth]{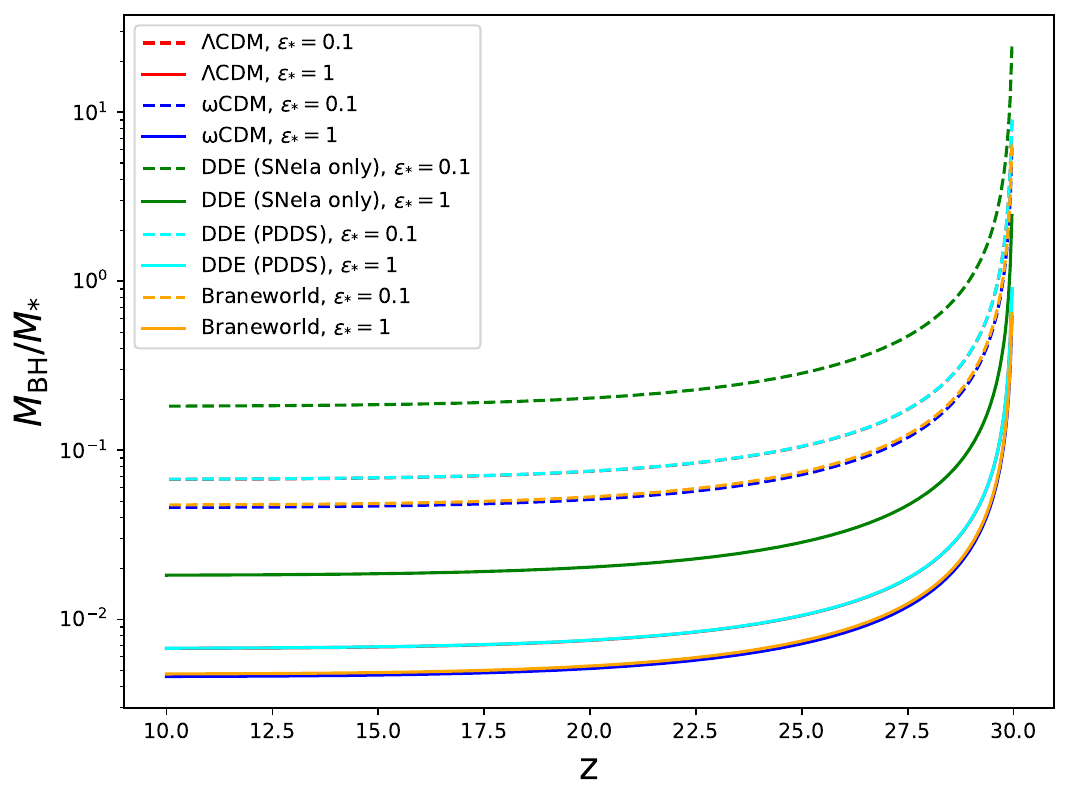}
		\caption{}
	\end{subfigure}
	\caption{Evolution of central black hole to stellar mass ratio ($M_{BH}/M_{*}$) with redshift for (a) $M_{BH}=10^{8} M_{\odot}$ and (b) $M_{BH}=10^{9} M_{\odot}$ in different cosmologies for SFE($\epsilon_{*}$) = 0.1 and 1. }
\end{figure}

\section*{Acknowledgements}

The corresponding author greatly acknowledges the Department of Science and Technology (DST), Govt. of India, for providing financial assistance through the WISE-PhD fellowship bearing the grant no. DST/WISE-PhD/PM/2024/24.

\section*{Data Availability}

All data underlying this study are included within the figures and tables in the article.



\bibliographystyle{mnras}
\bibliography{document} 





\label{lastpage}
\end{document}